\documentclass[iop]{emulateapj}
\newcommand{\Ser}{S\' ersic\ }
\def\ln{${\rm L_{\rm X,nuc}}\,\,$}

\def\mb{$M_{\rm BH}\,\,$}
\def\ba{\hskip-.1truecm}
\def\lb{${\rm L_{B}}\,\,$}
\def\ledd{${\rm L_{Edd}}\,\,$}
\def\md{$\dot M_{\rm BH}$}

\catcode`\@=11
\def\gsim{\ifmmode{\mathrel{\mathpalette\@versim>}}
    \else{$\mathrel{\mathpalette\@versim>}$}\fi}
\def\lsim{\ifmmode{\mathrel{\mathpalette\@versim<}}
    \else{$\mathrel{\mathpalette\@versim<}$}\fi}
\def\@versim#1#2{\lower 2.9truept \vbox{\baselineskip 0pt \lineskip
    0.5truept \ialign{$\m@th#1\hfil##\hfil$\crcr#2\crcr\sim\crcr}}}
\catcode`\@=12

\shortauthors{S. Pellegrini}
\shorttitle{The nuclear X-ray emission of early type galaxies}

\begin{document}

\title{The nuclear X-ray emission of nearby early-type galaxies}

   \author{S. Pellegrini\footnote{E-mail:  silvia.pellegrini@unibo.it}}
\affil{Astronomy Department, University of Bologna, 
                       via Ranzani 1, 40127 Bologna, Italy   }

\begin{abstract} 

Nuclear hard X-ray luminosities (\ln\ba) for a sample of 112 early
type galaxies within a distance of $67$ Mpc are used to investigate
their relationship with the central galactic black hole mass \mb
(coming from direct dynamical studies or the \mb$-\sigma$ relation),
the inner galactic structure (using the parameters describing its
cuspiness), the hot gas content and the core radio luminosity. For
this sample, \ln ranges from $10^{38}$ to $10^{42}$ erg s$^{-1}$, and
the Eddington ratio \ln\ba/\ledd from $10^{-9}$ to $10^{-4}$, with the
largest values belonging to four Seyfert galaxies. Together with a
trend for \ln to increase on average with the galactic luminosity \lb
and \mb\ba, there is a wide variation of \ln (and \ln\ba/\ledd\ba), by
up to 4 orders of magnitude, at any fixed \lb$> 6\times
10^{9}L_{B,\odot}$ or \mb$>10^7M_{\odot}$.  This large observed range
should reflect a large variation of the mass accretion rate \md, and
possible reasons for this difference are searched for. On the
circumnuclear scale, in a scenario where accretion is (quasi) steady,
\md$\,$ at fixed \lb (or \mb\ba) could vary due to differences in the
fuel production rate from stellar mass return linked to the inner
galactic structure; a trend of \ln with cuspiness is not
observed, though, while a tendency for \ln\ba/\ledd to be larger in
cuspier galaxies is present.  In fact, \md$\,$ is predicted to vary
with cuspiness by a factor exceeding a few only in hot gas poor
galaxies and for large differences in the core radius; for a subsample
with these characteristics the expected effect seems to be present in
the observed \ln values. \ln does not show a dependence on the
age of the stellar population in the central galactic region, for
ages$>$3 Gyr; less luminous nuclei, though, are found among the
youngest galaxies or galaxies with a younger stellar component.  On
the global galactic scale, \ln shows a trend with the total galactic
hot gas cooling rate ($L_{X,ISM}$): it is detected both in gas poor
and gas rich galaxies, and on average increases with $L_{X,ISM}$, but
again with a large scatter.  The observed lack of a tight relationship
between \ln and the circumnuclear and total gas content can be
explained if accretion is regulated by factors overcoming the
importance of fuel availability, as 1) the gas is heated by black hole
feedback and \md$\,$ varies due to an activity cycle, and 2) the mass
effectively accreted by the black hole can be largely reduced with
respect to that entering the circumnuclear region, as in radiatively
inefficient accretion with winds/outflows.  Finally, differently from
\ln\ba, the central 5 GHz VLA luminosity shows a clear trend with the inner
galactic structure, that is similar to that shown by the total soft
X-ray emission; therefore it is suggested that they could both be
produced by the hot gas.

\end{abstract}


\section{Introduction}\label{intro} 

In the past years, high angular resolution studies of the centers of
early type galaxies have been performed with the $Hubble$ $Space$
$Telescope$ ($HST$) in the optical and near infrared, and in the
X-rays with $Chandra$, obtaining important results that deeply
influenced our understanding of the nature and past evolution of these
systems.  The first, major $HST$ result was that massive black holes
(MBHs) are ubiquitous in the centers of spheroids, and linked by tight
relationships with the luminosity and central stellar velocity
dispersion of their hosts (e.g., Magorrian et al. 1998, Ferrarese \&
Merritt 2000, Gebhardt et al. 2000), indicative of a strong mutual
influence during their formation and evolution. The second $HST$
result was that the central brightness profiles of galaxies with
$M_V\lsim -19$ show either steep brightness cusps or, interior to a
break radius $r_b$, they flatten markedly in a core with respect to an
inner extrapolation of the outer profile.  These profiles have been
described respectively by a \Ser or core-\Ser law (Graham et
al. 2003, Trujillo et al. 2004, Ferrarese et al. 2006, Kormendy et
al. 2009) or alternatively by the "Nuker law" (Faber et al. 1997,
Lauer et al. 2007a). Cores dominate at the highest luminosities and
steep cusps at the lowest, with an intermediate luminosity region of
coexistence ($-20.5\gsim M_V \gsim -23$, Lauer et al. 2007a).  The
shape of the brightness profile in the inner galactic region has been
related to the past formation and evolution of galaxies, with cores
created during dry merging events by a black hole binary ejecting
stars from the center of the new system (Ebisuzaki et al. 1991, Faber
et al. 1997, Milosavljevic et al. 2002, Graham 2004, Gualandris \&
Merritt 2008), and cusps being preserved or (re)generated during
gaseous (wet) mergings.  Recently, it was found that coreless
elliptical galaxies in the Virgo cluster have extra-light at their
center, above the inward extrapolation of their outer \Ser profile
(Kormendy et al. 2009), the result of a wet merger induced starburst
(see also Hopkins et al. 2009a) or of AGN induced starburst activity
(Ciotti \& Ostriker 2007).  Moreover, the presence of steep cusps or
cores correlates with other fundamental galactic properties, even more
tightly than how these properties correlate with the galatic
luminosity: core galaxies generally have boxy isophotes, are slow
rotators and triaxial systems, while cusp galaxies are disky, fast
rotators and axisymmetric (Kormendy \& Bender 1996, Faber et
al. 1997); core galaxies show a large range of radio and X-ray
luminosities, while cusp galaxies are confined below a threshold
(Bender et al. 1989; Pellegrini 1999, 2005a; Capetti \& Balmaverde
2005; Pasquali et al. 2007).  Pellegrini (2005a) also attempted an
investigation of the relation between the X-ray nuclear emission (\ln)
and the inner core/cusp profile, but the study was limited by the
small number of nuclei with known \ln available.

Launched in 1999, the $Chandra$ satellite has now pointed a large
number of early type galaxies, with an unprecedented angular
resolution in the X-rays of less than $1^{\prime\prime}$.  For the
first time measurements of the nuclear X-ray emission down to values as
low as $10^{39}$ erg s$^{-1}$ and out to distances of $\sim 60$ Mpc
have been obtained.  The MBHs of the local universe turned out to be
typically radiatively quiescent and very sub-Eddington emitters
(Loewenstein et al. 2001, Soria et al. 2006a, Zhang et al. 2009,
Gallo et al. 2008, 2010).  
In a number of cases the mass accretion rate on the MBH
could be estimated (e.g., Di Matteo et al. 2003, Pellegrini 2005b) and
the radiative quiescence was interpreted in terms of radiatively
inefficient accretion (Narayan \& Yi 1995), possibly with the
mechanical power dominating the total output of accretion (e.g., Allen
et al. 2006).  However, many aspects of accretion in the local
universe remain unknown: what determines \ln ?  Is there any relation
of \ln or its Eddington-scaled value with the galactic luminosity or
the mass of the central supermassive black hole \mb\ba?  or with
the inner stellar profile, that has been linked to the past galactic
evolution and other major global galactic properties? or with the hot
gas content?  Answering these questions is important for a complete
understanding of the MBH-host galaxy coevolution process.

In this work we have collected all early type galaxies (E and S0) out
to a distance of $\sim 67$ Mpc with known \ln\ba, based mostly on data
coming from $Chandra$ pointings; a total of 112 galaxies resulted with
\ln measured or with an upper limit on it.  The sample includes also
most of the early type galaxies with a direct measurement of \mb via
dynamical studies currently available; for the other galaxies, the
central stellar velocity dispersion allows for an estimate of \mb via
the \mb$-\sigma$ relation.  For 81 of these 112 galaxies the central
stellar profile shape has been measured with $HST$.  The sample is
described in Sect.~\ref{sample}, the observational evidences about
relationships between \ln\ba, \mb\ba, the central stellar structure
and the radio luminosity are presented in Sect.~\ref{obs}, the results
are discussed in Sect.~\ref{disc} (that examines also the relationship
between \ln and the galactic hot gas luminosity); the conclusions are
summarized in Sect.~\ref{concl}.

\section{The sample}\label{sample}

The morphological type of the sample includes E and S0 objects, that
is early type galaxies with numerical code t$\leq -2$ according to the
revised de Vaucouleurs morphological type defined in RC2. A distance
limit of $\sim 70$ Mpc was set to allow for the possibility of a measurement
of \ln (or an upper limit on it) with $Chandra$ down to $\sim 10^{39}$
erg s$^{-1}$ even for the most distant objects, and of a measurement
of the inner light profile from $HST$ data for a large fraction of the
objects (most cores have a radius $<500$ pc). A selection of galaxies
with the chosen morphological properties and distance limit was
performed with the Hyperleda catalog\footnote{
http://leda.univ-lyon1.fr}. The resulting list was then
cross-correlated with the list of $Chandra$ pointings, using the Web
ChaSeR (Chandra Archive Search and Retrieval Interface\footnote{
http://asc.harvard.edu/cda/chaser.html}), to find the objects with
X-ray information on their nuclei. Published works as of December 2009
based on these pointings provide a detection or upper limit for the
nuclear luminosity \ln for 97 galaxies.  In order to avoid
possible contamination from soft hot gaseous emission, \ln was taken
in the 2--10 keV band, or converted to it based on the spectral shape
used to derive it.  The object list, with adopted distances, \ln
values and references for them, is given in Tab.~\ref{tab1}.

For the objects in this list, the light profile shape in the inner
regions was then searched.  Two central slopes have been considered
previously (Lauer et al. 1995, Rest et al. 2001, Lauer et
al. 2007a): $\gamma^{\prime}$, the local slope evaluated at the $HST$
angular resolution limit, and $\gamma$, the slope describing the
brightness profile $I(R)\propto R^{-\gamma}$ interior to a break
radius $r_b$, when adopting a ``Nuker law" profile description
(a broken power law with a smooth transition from the outer slope to
the inner slope $\gamma$).  Steep inner cusps have $\gamma$ and
$\gamma^{\prime}$ larger than 0.5, and their host galaxies are called
"power law" or "cusp" galaxies; cores have $\gamma $ and
$\gamma^{\prime}$ smaller than $0.3$ (with $\sim 10$\% of galaxies
with $\gamma< 0.3$ that have $\gamma^{\prime}>0.3$); ``intermediate''
systems are a minority and have $\gamma$ or $\gamma^{\prime}$ between
0.3 and 0.5.  In core galaxies a well defined break radius $r_b$ marks
a rapid transition from the outer profile to a much shallower inner
slope; cusp galaxies retain a steep slope into the resolution
limit. The definition of what constitutes a core (i.e., a break with
respect to a \Ser function fitted to the outer profile, or an inner
slope $\gamma <0.3$ in the Nuker function fit) is different for the
two descriptions using the two functions, but gives the same
classification as core or cusp galaxy for most galaxies (Kormendy et
al. 2009).  For most of the galaxies in table~\ref{tab1}, the inner
profile shape comes from the large compilation of Lauer et
al. (2007a), who combined the results of previous {\it HST}
investigations of the central structure of early-type galaxies, after
transformation to a common band and distance scale; for other 18
galaxies the Nuker law description of the $HST$ profile is taken from
Capetti \& Balmaverde (2005; these are marked in col. 5 in
Tab.~\ref{tab1}).  For the purpose of investigating the relationship
between \ln and the inner galactic structure with as many objects as
possible, the initial sample of 97 galaxies was enlarged with 15
objects with inner $HST$ slope measured. Of these, 7 follow all the
adopted selection criteria, but their \ln comes from $ROSAT$ HRI
pointings (with an angular resolution of $\sim 5^{\prime\prime}$, from
Liu \& Bregman 2005); the other 8 galaxies, with \ln from $Chandra$
data, have a type later than t=-2, though still within the S0
range\footnote{These 8 galaxies are NGC7743 (t=-0.9), NGC524, NGC3945
(t=-1.2), NGC4382, NGC5866 (t=-1.3), NGC4459, NGC4111 (t=-1.4),
NGC1316 (t=-1.7). For reference, type S0a has t=0.}.  Tab.~\ref{tab1}
then includes 112 galaxies, 47 of core type, 7
intermediate, and 27 of cusp type.  The profile classification (cusp or
core) is based on the value of $\gamma$, given in tab.~\ref{tab1};
this classification is always coincident with that given by
$\gamma^{\prime}$, except for 7 cases that become "cuspier" when
considering $\gamma^{\prime}$ instead of $\gamma$\footnote{These are 5
cores and one intermediate that become cusps, and one core that
becomes intermediate; all these cases are marked in the following
figures with a special symbol, as explained in the captions. Note
that all galaxies with $\gamma $ in Tab.~\ref{tab1} have $M_V<-19$
(from the apparent magnitudes in Hyperleda, and for the distances adopted
in Tab.~\ref{tab1}), a luminosity range where low $\gamma<0.3$ values
are associated only to cores in massive ellipticals,
and high $\gamma$'s unequivocally identify coreless galaxies
 (Trujillo et
al. 2004, Lauer et al. 2007a).}. The results of this work are
unchanged when using $\gamma^{\prime}$ instead of $\gamma$.  Other
quantities used in the following and not given in Tab.~\ref{tab1} are
the break radii $r_b$ and effective radii $R_e$; these are taken from
Lauer et al. (2007a) and Capetti \& Balmaverde (2005); $r_b$ ranges
from a few parsecs to few hundreds of parsecs.

Finally, Tab.~\ref{tab1} lists the masses of the central MBHs, taken
from specific measurements for a number of galaxies, and from the
$M_{BH}-\sigma $ relation for the other cases, as specified in the
table.  The adopted $M_{BH}-\sigma $ relation is that recently derived
for ellipticals (G\"ultekin et al. 2009), in which the central stellar
velocity dispersion $\sigma$ from the Hyperleda catalog is inserted; this
relation has an intrinsic scatter of 0.31 dex in log\mb\ba. Note that \mb
values estimated from the $M_{BH}-L_V $ relation of early type
galaxies (where $L_V$ is the V-band luminosity; G\"ultekin et
al. 2009) can be larger than those derived from the $M_{BH}-\sigma $
relation, for \mb$\lsim 10^7M_{\odot}$ [see also Gallo et al. (2008)].

The galaxies in Tab.~\ref{tab1} reside in all types of environment,
going from being isolated to being at the center of a cluster (as
NGC4486 in Virgo) or a group. They also span a large range of
activity, from being classified as an optical or radio AGN (e.g.,
Seyfert or FRI) to inactive.  Their optical nuclear spectra have been
classified as absorption nuclei (i.e., without emission lines) or,
when emission lines are present, mostly as LINERs, with a minority of
Seyfert and transition nuclei (intermediate between HII and LINERs; Ho
et al. 1997).

\section{Observational results}\label{obs}

Figure~\ref{f1} shows the relationship between \ln and the B-band
galactic luminosity \lb\ba.  Considering the whole \lb range, \ln
increases on average with \lb\ba.  However, in the well populated
region of the plot, for \lb$> 6\times 10^{9}L_{B\odot}$, \ln does not
present a clear trend with \lb but shows instead a very large scatter
of $\sim 4$ orders of magnitude, going from the lowest detectable \ln
values in the nearest galaxies ($\sim 10^{38}$ erg s$^{-1}$) to the
highest \ln values ($\gsim 10^{42}$ erg s$^{-1}$) that belong to
Seyfert nuclei (NGC2110, NGC3516, NGC5128, NGC5283; all these are S0
galaxies, consistent with the observation that Seyfert nuclei reside
mostly in spirals and S0s, e.g., Malkan et al. 1998).  Similar
considerations hold for fig.~\ref{f2}, where \lb is replaced by
\mb\ba, as expected given the Magorrian relation between the spheroid
luminosity and the MBH mass (Magorrian et al. 1998): there is an
overall trend for \ln to increase with \mb\ba, but in the well
populated region, for \mb$>2\times 10^7$ M$_{\odot}$, \ln shows a very
large scatter, of $\sim 4$ orders of magnitude.  Figures~\ref{f1} and
\ref{f2} show cusp and core galaxies being more frequent respectively
at the lowest and highest \lb (and \mb\ba) values, but with a large
intermediate region of overlap; more importantly, these figures also
show that both types cover the whole large range of \ln\ba.

Figure~\ref{f3} shows the relation between \ln scaled by the Eddington
luminosity [i.e., \ln\ba/\ledd\ba, where \ledd\ba=$1.25\times
10^{38}$\mb\ba(M$_{\odot}$) erg s$^{-1}$] and \mb\ba. Since \ledd is
proportional to \mb\ba, the lower envelope of the \ln values in
fig.~\ref{f2} (a straight horizontal line at \ln$\sim 10^{38}$ erg
s$^{-1}$) translates in a lower envelope of \ln\ba/\ledd values that
is a line with slope -1, shown in fig.~\ref{f3} as a dotted line.
Above this line, a large scatter of $\sim 3-4$ orders of magnitude in
\ln\ba/\ledd is evident also in this plot. Again, both core and cusp
galaxies cover a large range of \ln\ba/\ledd\ba; in this case,
though, the "cuspier" types seem to reach highest Eddington ratios of
$10^{-4}-10^{-3}$ and the core types to be confined below
\ln\ba/\ledd\ba$<10^{-4.5}$. Except for few very nearby galaxies,
Eddington ratios similarly low cannot be reached at the lowest and at
the highest \mb values, due to the limit marked by the dotted line;
but the highest Eddington ratios, for which there are no limits, tend
to correspond to the lower \mb\ba. A similar result was obtained for
nearby late type galaxies (Zhang et al. 2009) and for the 100
spheroidal galaxies of the ACS Virgo cluster survey studied 
by Gallo et al. (2010).
In the latter survey, \ln was estimated with $Chandra$ down
to a limit of L(0.5-7 keV)$=3.7\times 10^{38}$ erg s$^{-1}$ for a
sample with stellar masses peaking below $10^{10}M_{\odot}$ and MBH
masses below $10^8M_{\odot}$; the Eddington ratio
was found to decrease on average as \mb\ba$^{-0.62}$, with a
scatter of 0.46 dex. This trend is consistent with the
distribution of the \ln values in fig.~\ref{f3}, though the large
scatter here dominates.  The average decrease of the Eddington ratio
with increasing \mb was interpreted as a manifestation of down-sizing
in black hole accretion (Gallo et al. 2010, Schawinski et al. 2010).

\begin{figure*}
\hskip 0.8truecm
\includegraphics[height=0.8\textheight,width=1\textwidth]{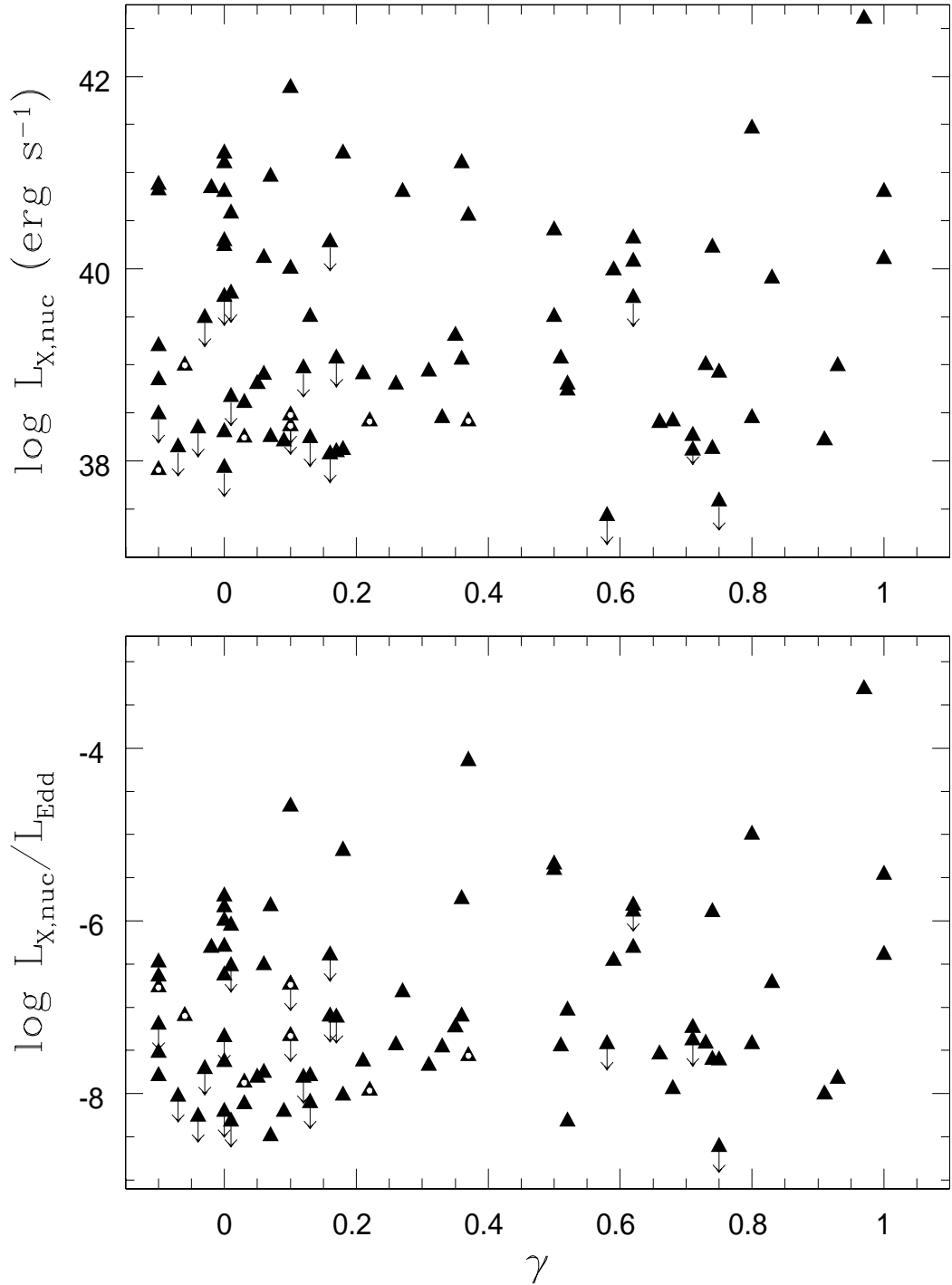}
\caption{The relationships between \ln and the slope of the
stellar light profile in the
central galactic region $\gamma$ (upper panel), and between \ln\ba/\ledd
and $\gamma$ (lower panel). Upper limits on \ln are shown with an arrow.
Galaxies that change their classification from core to cusp or intermediate
when using $\gamma^{\prime}$ instead f $\gamma$ 
(Sect.~\ref{sample}) are marked with a white dot inside their symbol.}
\label{f4}
\end{figure*}

\begin{figure*}
\hskip 2truecm
\vskip -8truecm
\includegraphics[height=0.8\textheight,width=1\textwidth]{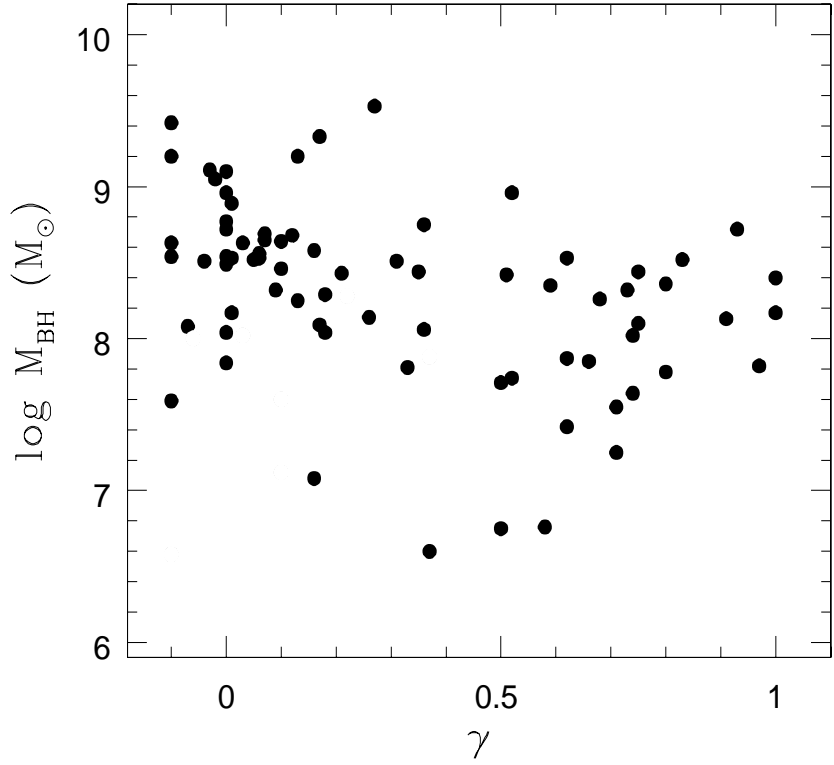}
\caption{The relationships between the mass of the MBH and the slope of the
stellar light profile in the
central galactic region $\gamma$ for the sample in tab.~\ref{tab1};
the 7 galaxies that change their classification from core to cusp or 
intermediate
when using $\gamma^{\prime}$ instead of $\gamma$
(Sect.~\ref{sample}) have been excluded.}
\label{f5}
\end{figure*}

In order to better evidence any possible relationship between \ln and
the inner light profile, \ln and \ln\ba/\ledd were plotted against the
slope of the central light profile $\gamma$ (fig.~\ref{f4}).  As
already suggested by figs.~\ref{f1} and \ref{f2}, fig.~\ref{f4} shows
that \ln spans the whole large range of values, from $\sim 10^{38}$ to
$\gsim 10^{42}$ erg s$^{-1}$, at all $\gamma$'s. A similar result
holds for \ln\ba/\ledd\ba (fig.~\ref{f4}, lower panel), though
here the highest \ln\ba/\ledd values among core galaxies remain lower
than those of intermediate and cusp galaxies, as seen from
fig.~\ref{f3}.  This possible increase of the upper envelope of values
of \ln\ba/\ledd with $\gamma$, though requiring a larger and complete
sample for a firm conclusion, may be real, given that the galaxies considered
here already include more core than cusp cases, and that such an
increase is expected to arise, when considering the trend for the
Eddington ratio to increase for \mb decreasing (fig.~\ref{f3}), and
the trend for \mb to decrease on average with $\gamma$ increasing (as
shown in fig.~\ref{f5} for the present sample).  This latter behavior
is a consequence of the Magorrian \mb-$L_B$ relation coupled with the
weak $\gamma -L_B$ anti-correlation (Faber et al. 1997 and
Sect.~\ref{intro}).  Another possible reason for an increase of
\ln\ba/\ledd with $\gamma$ is discussed in Sect.~\ref{mrate}.

Finally, to explore further a possible link with the inner galactic
structure, we also considered the relation of \ln with the break
radius $r_b$, and with $r_b$ rescaled by $R_e$.  Only core galaxies
were considered, for which $r_b$ is a representative scale for the
radial extent of the core (Lauer et al. 2007b). When plotted against
$r_b$, \ln covers the same large range of values differing by $\sim 3$
orders of magnitude already seen in the previous figures, and shows no
trend with $r_b$ or $r_b/R_e$.  The same holds for \ln\ba/\ledd\ba.

\subsection{The different relation with cuspiness
of \ln and the radio luminosity}\label{xr}

As mentioned in Sect.~\ref{intro}, for a sample of galaxies with
central light profiles measured from $HST$ data, and with 5 GHz radio
fluxes estimated from a VLA survey with a $\sim 3-5^{\prime\prime}$
FWHM resolution, Capetti \& Balmaverde (2005) found the radio emission
of core galaxies to cover a large range of values, differing by orders
of magnitude, while that of cusp galaxies to be confined below
$L_R\sim 3\times 10^{21}$ W Hz$^{-1}$. The origin of this threshold in
radio luminosity for cusp galaxies remained ambiguous, because
galaxies with a higher intrinsic optical luminosity have a
higher probability to be strong radio emitters, with respect to less
luminous galaxies (e.g., Mauch \& Sadler 2007). This is true also
for the Capetti \& Balmaverde (2005) sample, whose $L_R - M_K$
relation (where $M_K$ is the K-band absolute magnitude) shows no radio
source with $L_R$ above the threshold of $3\times 10^{21}$ W Hz$^{-1}$
associated to a host with $M_K>-24$. Since there are only few cusp
galaxies in their sample with $M_K<-24$, it cannot be concluded
whether the threshold is related to a different nuclear structure, or
to cusp galaxies populating scarcely the $M_K$ range where the
brightest radio sources are found.

Both the radio and the X-ray emission are signatures of MBH accretion,
but fig.~\ref{f4} (upper panel) shows clearly the lack of a threshols for
\ln similar to that found for $L_R$; therefore, the behavior of \ln
and $L_R$ with respect to the presence of a core in the stellar light
profile is here revisited.  Figure~\ref{f4} was then remade just for
the Capetti \& Balmaverde (2005) sample, that includes 51 objects, 42
of which are in Tab.~\ref{tab1}; all these 42 galaxies have
$M_B<-18.6$, a luminosity range where low $\gamma<0.3$ values are
associated only to cores in massive ellipticals (Trujillo et
al. 2004). For this sample, figure~\ref{f6} shows the relationship
between $\gamma$ and the 5 GHz core luminosity $L_{R,core}$ from the
VLA survey (taken from Capetti \& Balmaverde 2006 or Balmaverde \&
Capetti 2006, and rescaled for the distances in tab.~\ref{tab1}).  The
figure shows again a different behavior with respect to $\gamma$ of
\ln and of $L_{R,core}$: while there is an L-shape distribution of
$L_{R,core}$ with respect to the inner light profile, with galaxies of
all types below $L_{R,core}\sim 10^{38}$ erg s$^{-1}$ and only core
galaxies above, \ln shows the same lack of a threshold found in
fig.~\ref{f4}.  We note that the VLA radio data 
used to derive the luminosities in fig.~\ref{f6} do not
separate well the core emission from any extended structure, and they
tend to overestimate the core flux when there is extended radio emission
(Balmaverde \& Capetti 2006).  These findings
are discussed in Sect.~\ref{difxr}.

\section{Discussion}\label{disc}

Hard X-ray emission is a major signature of accretion on a MBH, both
in the standard disk plus hot corona modality (Haardt \& Maraschi
1993) and in the radiatively inefficient modality (RIAF, Narayan \& Yi
1995) that is expected to establish at the low Eddington ratios of the
sample investigated here (fig.~\ref{f3}).  In fact, the nuclei in this
sample are highly sub-Eddington for any bolometric correction
$L_{bol}/$\ln that can be plausibly adopted: from $\sim 10$ for low
luminosity AGNs, as indicated by observations (e.g., Ho 2008) and
by the RIAF modeling (Mahadevan 1997), up to 40--70 for standard,
bright AGNs (Vasudevan \& Fabian 2007).  For radiatively
inefficient accretion \ln should scale inversely proportional to \mb\ba,
at fixed mass accretion rate on the MBH, \md 
(Mahadevan 1997; see also K\"ording et al. 2006).  What instead
dominates in fig.~\ref{f1} is a wide variation of \ln by 3--4 orders
of magnitude at any fixed \mb$>10^7 M_{\odot}$.  A large variation of
\ln of a similar extent was obtained also for other local samples,
made of galaxies of types later
than that discussed here and residing within 15 Mpc (Zhang et
al. 2009), or for the nuclei in the Palomar survey (Ho 2009).

Intrinsic variability of \ln cannot account entirely for the large spread in
figs.~\ref{f1}--\ref{f3}, since amplitude variations of low luminosity
AGNs are typically small (Ptak et al. 1998) and the largest keep within
a factor of a few (Pian et al. 2009).  The large range of \ln and of
\ln\ba/\ledd means then that physical quantities other than the size
of \mb play an important role. In the RIAF models a number of
parameters describe the complex physics of accretion (e.g., the ratio
of gas to magnetic pressure, the viscosity, the turbulent energy that
heats the electrons), but their variation is not expected to account
for \ln differences as large as those in figs.~\ref{f1}--\ref{f3}
(e.g., Di Matteo et al. 2003). Another major parameter determining \ln is
the mass accretion rate on the MBH, \md; for example, at fixed
\mb\ba, \ln scales as $\dot M_{\rm BH}^{2}$ for radiatively
inefficient flows (Narayan \& Yi 1995).  Below we discuss possible
variations of \md$\,$ produced by the inner galactic structure, that
determines how much mass is shed by stars in the circumnuclear region
and then the fuel available for the MBH.  In fact, the present sample
is defined morphologically to include galaxies with a typically old
stellar population, and in these systems the black hole growth is
regulated by the rate at which evolved stars lose their mass, as shown
for a large sample of nearby galaxies drawn from the Sloan Digital Sky
Survey (Kauffmann \& Heckman 2009).  In the following
Section~\ref{mrate} we assume that accretion is a (quasi) steady
process; then in Section~\ref{mdot} we consider the possibility that
accretion is unsteady.

\subsection{The mass accretion rate, the inner galactic structure
and the total gas content}\label{mrate}

Numerical simulations of the collective evolution of stellar mass
losses (Parriott \& Bregman 2008), for stellar mass distributions as
steep as revealed by $HST$ in the central galactic region, showed that
they establish an inflow towards the MBH (Pellegrini \& Ciotti 1998,
Pellegrini et al. 2007a). The size of the inflowing region $r_{in}$
can range from being of the order of the MBH accretion radius $r_a$
($r_a\propto $\mb\ba$/T_{ISM}= 10-100$ pc for typical values of the
ISM temperature $T_{ISM}$ and \mb=$10^8-10^9M_{\odot}$, e.g., Soria et
al. 2006a) to few kpc.  The stellar mass losses within $r_{in}$ then
make up the fuel to which \md$\,$ should be proportional. Since the
rate at which a given volume of a galaxy is replenished by mass losses
from stars during their passive evolution is just proportional to the
local stellar luminosity (e.g., Ciotti et al. 1991, David et
al. 1991), \md$\,$ and then \ln should depend on the shape of the
stellar density within $r_{in}$.  On average, core galaxies have a
lower central surface brightness and central stellar density than cusp
ones (Faber et al. 1997, Gebhardt et al. 1996, 2003, Kormendy et al.
2009), and core galaxies with larger cores have a lower central
density than those with smaller cores.  This determines that, at
similar \lb (or equivalently \mb\ba), cusp galaxies produce locally
more fuel for accretion than core ones, and galaxies with smaller
$r_b$ produce more fuel than those with larger $r_b$, which should
correspond to a higher \ln and \ln\ba/\ledd\footnote{Note that the
mass within the central few hundreds of pc is a small fraction ($<
0.1$) of the total stellar mass (Kormendy et al. 2009), so that
galaxies may have a very similar total luminosity or \mb (e.g., in
figs.~\ref{f1} and \ref{f2}) but different masses at the center.}. To
quantify this difference, we can calculate for example how the stellar
mass within three fiducial radii of $r_{in}=10$, 100 and $10^3$ pc,
representative of $r_a$ and of larger inflowing regions, varies
between a pure \Ser profile and a core-\Ser one that is identical to
the pure \Ser except for an inner slope $\gamma =0.1$ within a break
radius $r_b $ (e.g., Graham et al. 2003). We choose two
effective radii ($R_e=8$ or 12 kpc) and \Ser indeces ($n=6$ or $n=8$)
appropriate for core galaxies of \lb=2 or $5\times 10^{10}L_{B,\odot}$
(Ferrarese et al. 2006, Kormendy et al. 2009).  Within $r_{in}= 10$
pc, a pure \Ser profile has a mass 2--3 times larger than with
$r_b=10$ pc, and $\sim 25$ ($n=6$) or $\sim 50$ ($n=8$) times larger
than with $r_b=100$ pc.  Within $r_{in}= 100$ pc, a pure \Ser has
$\sim 3$ times more mass than a core-\Ser with $r_b=100$ pc, and $\sim
10$ ($n=6$) or $\sim 20$ ($n=8$) times more than for $r_b=300$ pc.
Within $r_{in}=1$ kpc, instead, a pure \Ser has a mass only $\lsim
20$\% larger than with $r_b=100$ or 300 pc, both for $n=6$ and
$n=8$. Therefore the difference in fuel production can be significant
(i.e., larger than a factor of a few) only for small inflow regions of
the order of $r_{in}\gsim r_a$, and large break radii $r_b> 10$ pc,
while it vanishes for $r_{in}\gsim 1$ kpc \footnote{According to the
most recent characterization of the central light profiles of early
type galaxies in Virgo, cusp galaxies are described by a \Ser profile
with $n<3$ plus an additional extra-light at the center with respect
to it (Kormendy et al. 2009). Therefore the above calculations give a
correct estimate for the variation of the central mass of core
galaxies with different $r_b$, but only an approximate estimate for
the variation between cusp and core galaxies.}.  Small $r_{in}$
correspond to low galactic hot gas contents, as found at \lb$\lsim
3\times 10^{10}L_{B,\odot}$ (David et al. 2006), and whenever the
total galactic soft X-ray emission $L_{X,tot}$ rescaled by \lb
(that is $L_{X,tot}$/\lb\ba) is low (Pellegrini \& Ciotti 1998); a difference
in \ln could then best manifest itself for these objects.  Similar
numbers to those obtained above for the stellar mass variation within
$r_{in}$ were obtained in the calculation of the stellar mass fraction
within different radii for the Lauer et al. (2007a) sample, divided in
bins of fixed total stellar mass; this fraction was found to show a
scatter of $\sim 1$ order of magnitude within $0.001R_e$ (Hopkins et
al. 2009b).  To summarize, 1) a difference in the inner galactic
structure can produce a different amount of circumnuclear material at
fixed \lb (or \mb\ba), and may then contribute to the scatter in
figs.~\ref{f1}--\ref{f3}, though it cannot account for the bulk of it,
since for most galaxies the variation in gas production within
$r_{in}$ will not exceed a factor of a few; 2) possible differences in
\md$\,$ are expected to exceed a factor of 2--3 only at
the lowest $L_{X,tot}$/\lb\ba, and should then be searched for in such
galaxies.

\begin{figure*}
\vskip -2truecm
\includegraphics[height=0.65\textheight,width=0.85\textwidth]{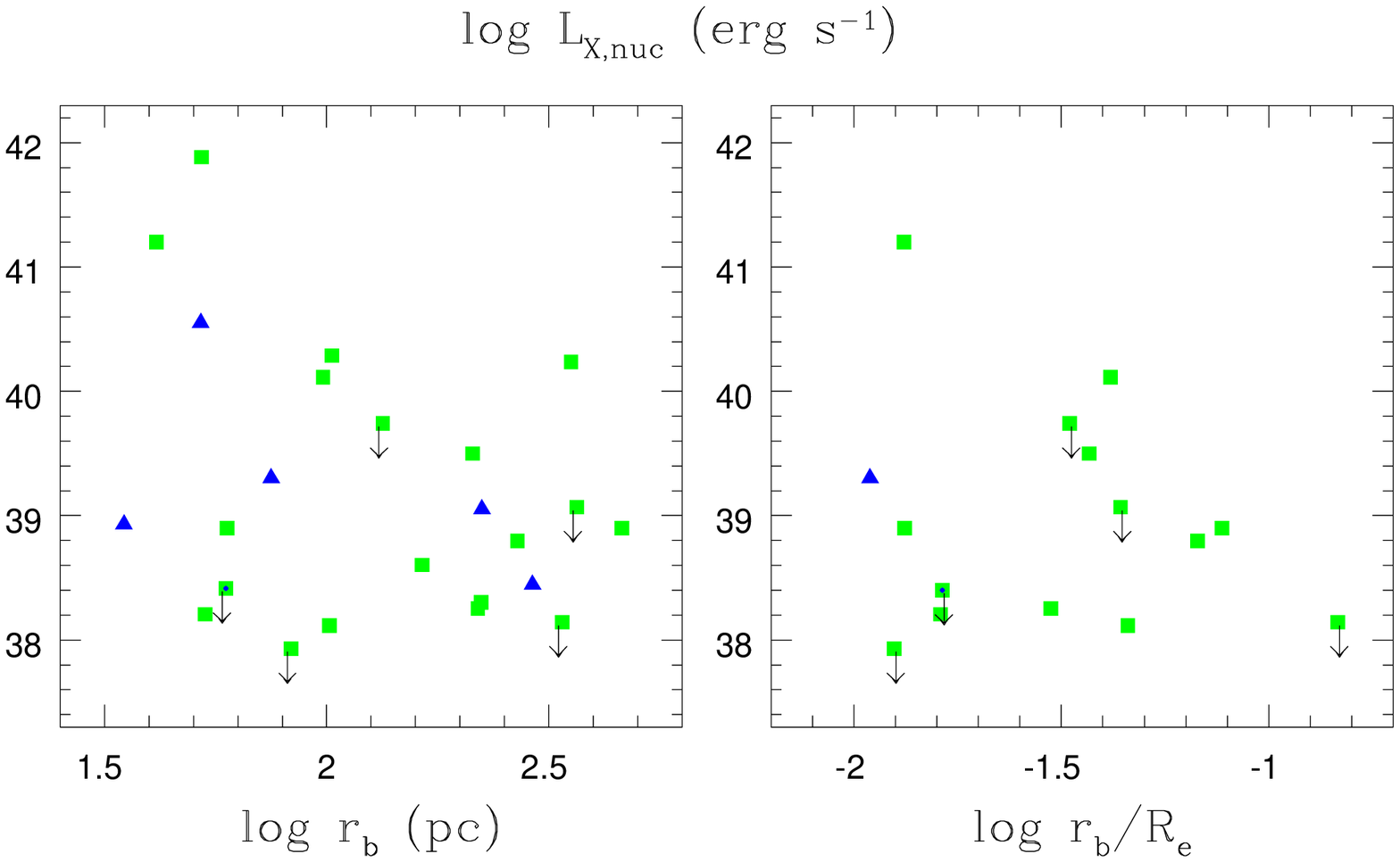}
\vskip -5truecm
\includegraphics[height=0.65\textheight,width=0.85\textwidth]{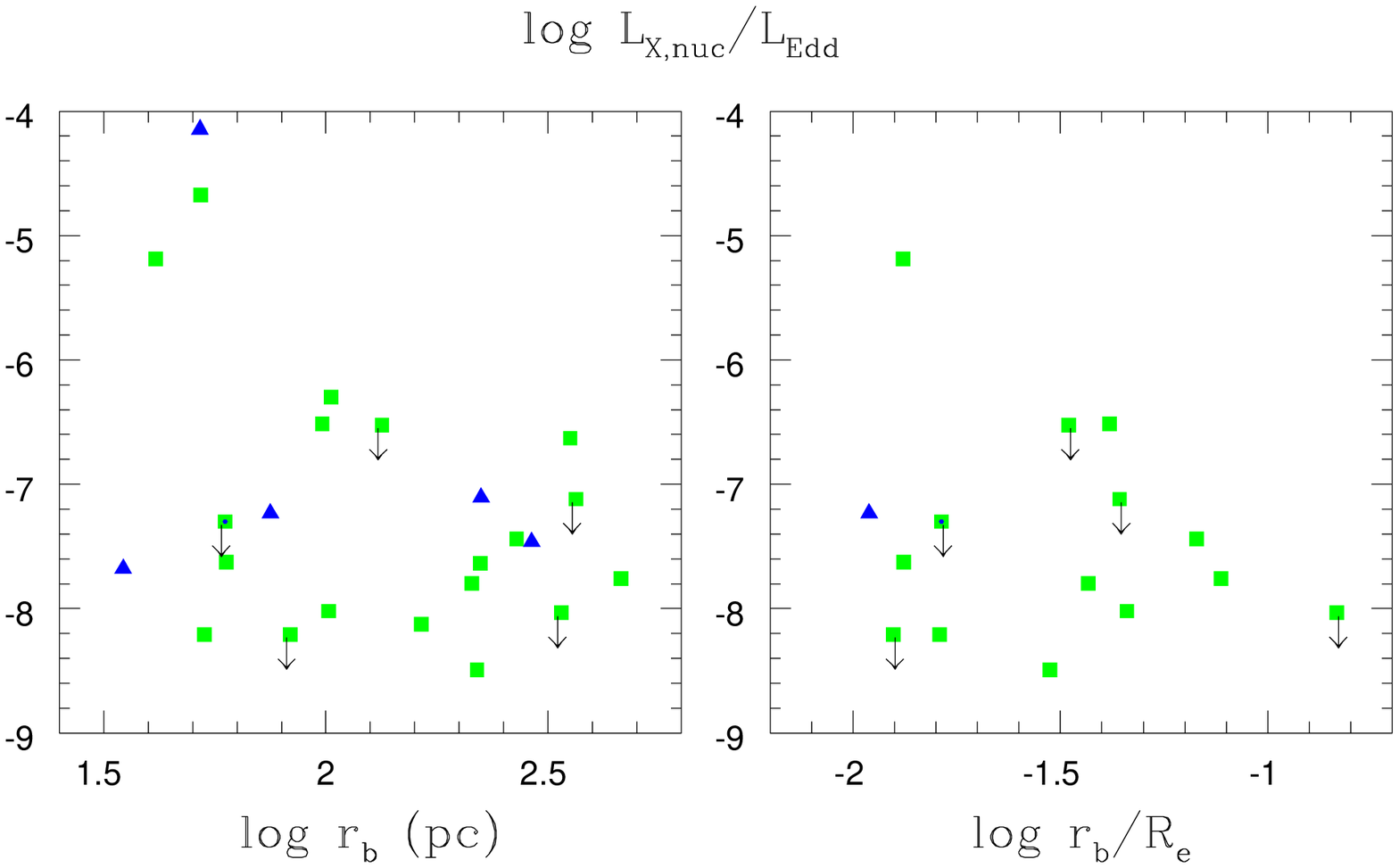}
\vskip -2truecm
\caption{\ln (upper panels) and \ln\ba/\ledd (lower panels) 
versus $r_b$ and $r_b/R_e$ 
for galaxies with a low hot gas content
(see Sect.~\ref{mrate}).}
\label{f7}
\end{figure*}

To pursue further point 2) above, we extracted from the list of
galaxies in tab.~\ref{tab1} a sample of core galaxies with low
$L_{X,tot}$/\lb\ba.  The total X-ray luminosity $L_{X,tot}$ was taken
from O'Sullivan et al. (2001), based on $ROSAT$ observations for a
thermal spectrum of $kT=1$ keV, and then most sensitive to the soft
gaseous emission; we considered the ratio $L_{X,tot}$/\lb to
correspond to a low hot gas content when $L_{X,tot}$(erg
s$^{-1}$)/\lb\ba$(L_{B,\odot}) \lsim 30.21$, that is below the upper
limit on the expected contribution from X-ray binaries (from Kim \&
Fabbiano 2004).  Figure~\ref{f7} shows \ln and \ln\ba/\ledd versus
$r_b$ and $r_b/R_e$ for this subsample: a trend in the predicted
direction is present for the upper envelope of the \ln and
\ln\ba/\ledd values, that decreases for increasing $r_b$.

\subsection{\ln and the age of the stellar population}\label{age}

\begin{figure*}
\vskip -5truecm
\hskip -1truecm
\includegraphics[height=1.1\textheight,width=1.4\textwidth]{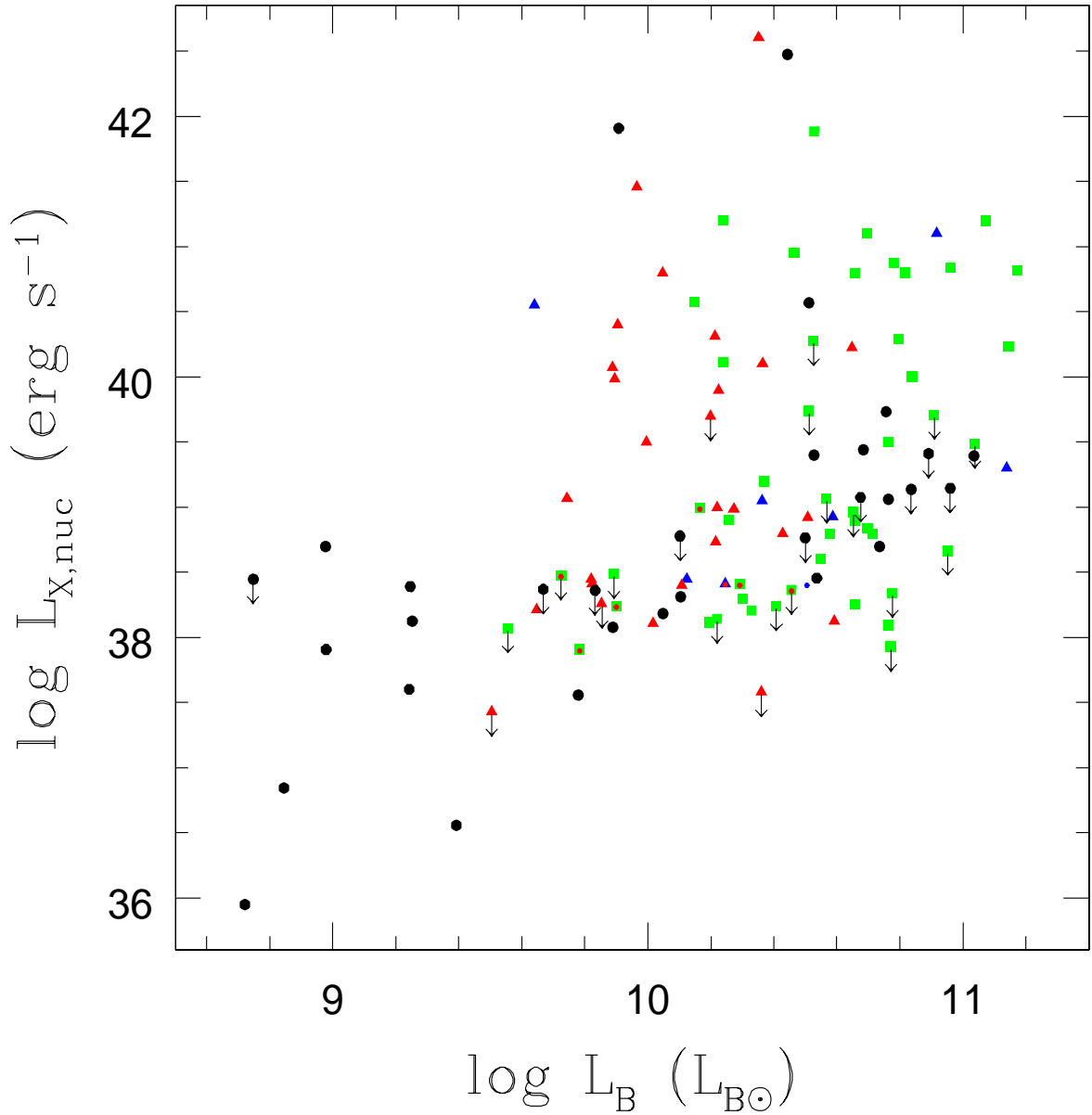}
\vskip -2truecm
\caption{The relation between the nuclear luminosity in the 2--10 keV band 
\ln and the galactic luminosity \lb for the sample described in 
Sect.~\ref{sample}. Upper limits on \ln are marked by
an arrow; red and blue triangles are cusp and intermediate galaxies,
green squares are core ones (all according to their $\gamma$ values
in Tab.~\ref{tab1}), 
unknown profiles are shown with black circles. Those 7 cases that become 
cusp (intermediate), when using $\gamma^{\prime}$ instead 
of $\gamma$ for their classification, are marked with 
a red (blue) dot inside the symbol (see Sect.~\ref{sample} and~\ref{obs}
 for more details). }
\label{f1}
\end{figure*}

\begin{figure*}
\vskip -5truecm
\hskip -1truecm
\includegraphics[height=1.1\textheight,width=1.4\textwidth]{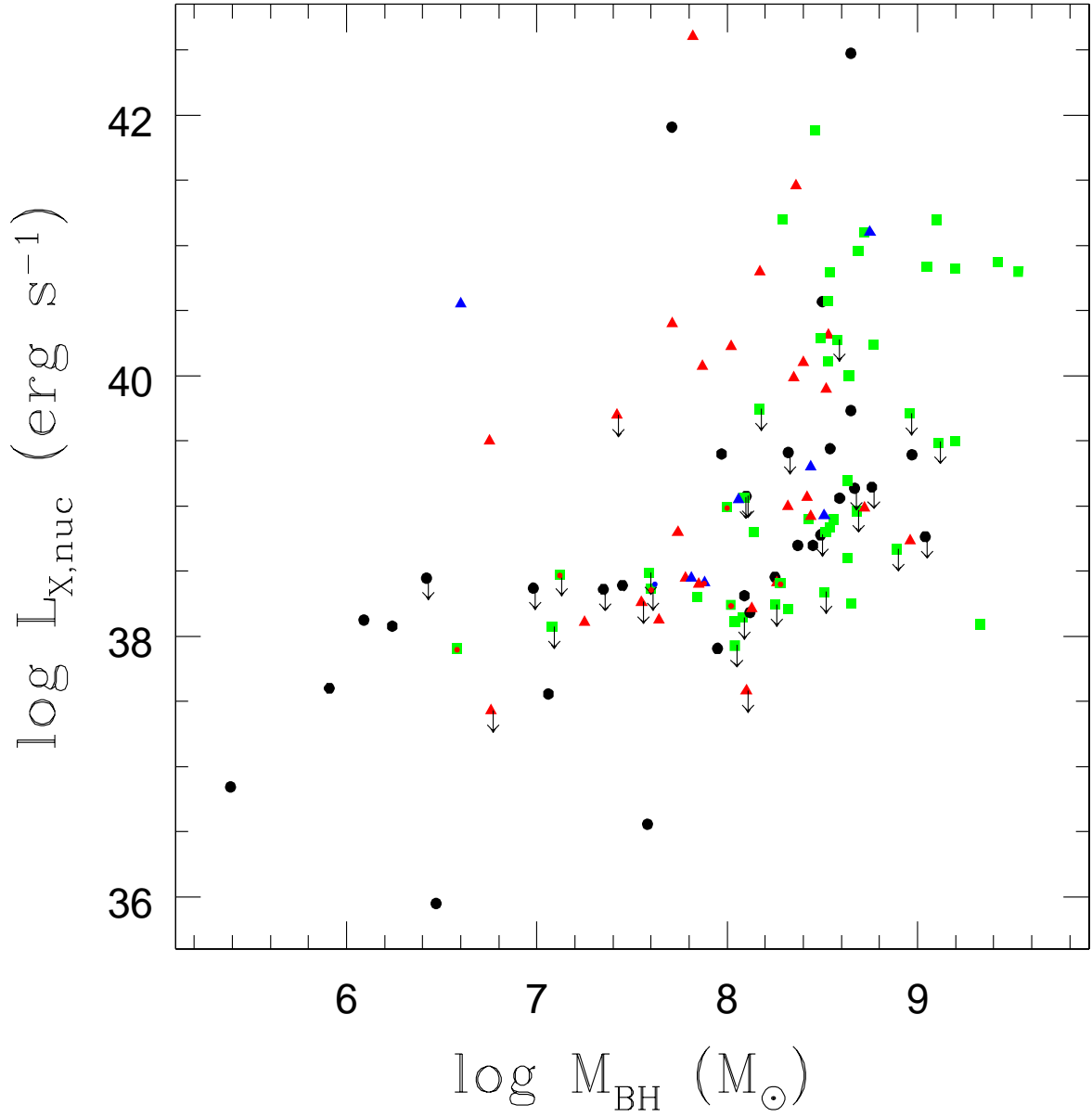}
\caption{The relation between the nuclear luminosity in the 2--10 keV band 
\ln and the MBH mass \mb\ba, with symbols as in fig.~\ref{f1};
\mb comes from direct estimates or the \mb$-\sigma$ relation,
as specified in Tab.~\ref{tab1}
(see Sects.~\ref{sample} and~\ref{obs} for more details).}
\label{f2}
\end{figure*}

\begin{figure*}
\vskip -7truecm
\includegraphics[height=1.1\textheight,width=1.4\textwidth]{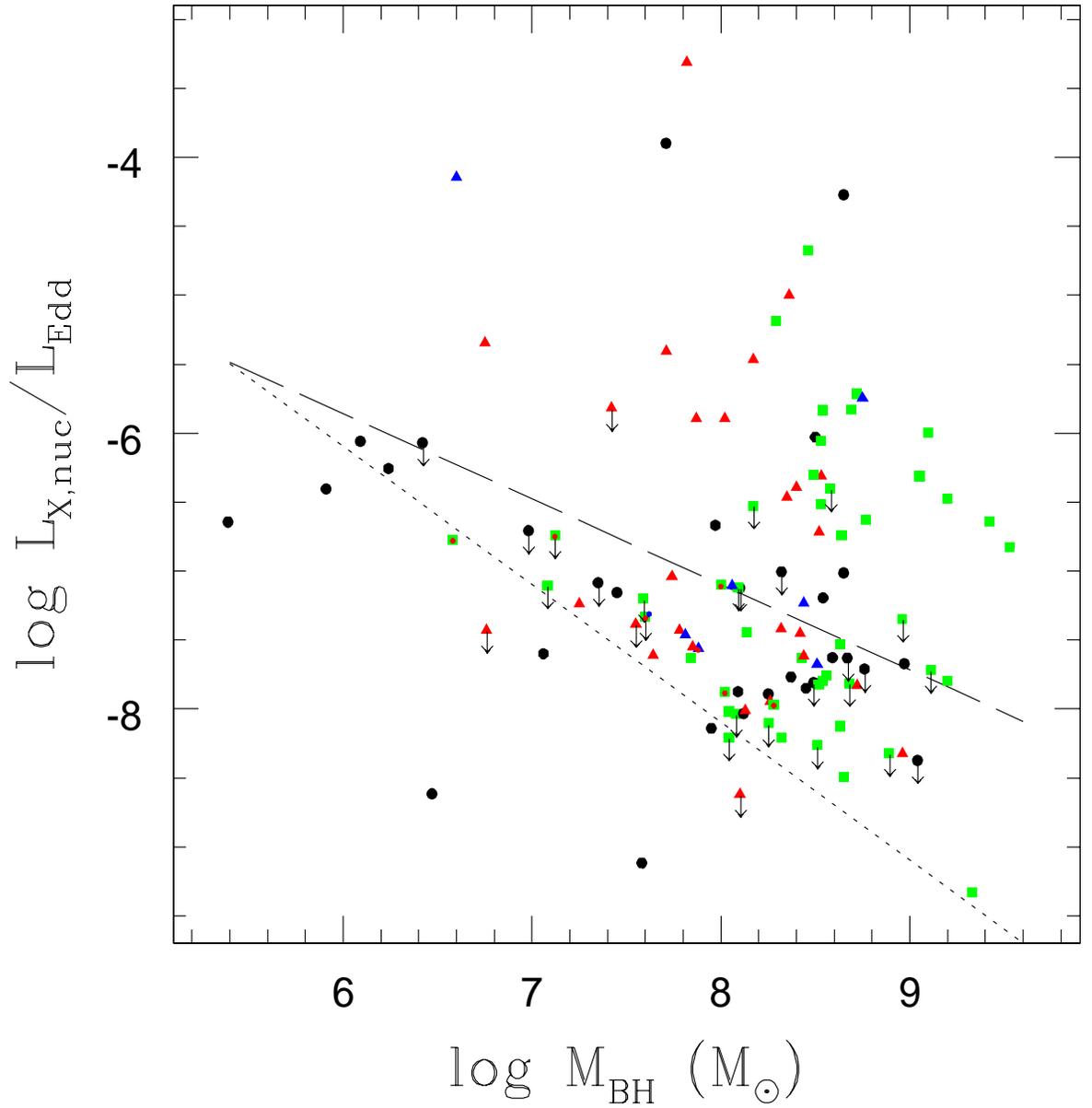}
\caption{The relation between \ln\ba/\ledd and \mb\ba, with symbols as 
in fig.~\ref{f1}; the dashed line is the average trend found for the 
ACS VCS sample by Gallo et al. (2010)
(see Sect.~\ref{obs} for more details).}
\label{f3}
\end{figure*}

\begin{figure*}
\vskip -6truecm
\hskip 1.truecm
\includegraphics[height=0.75\textheight,width=0.85\textwidth]{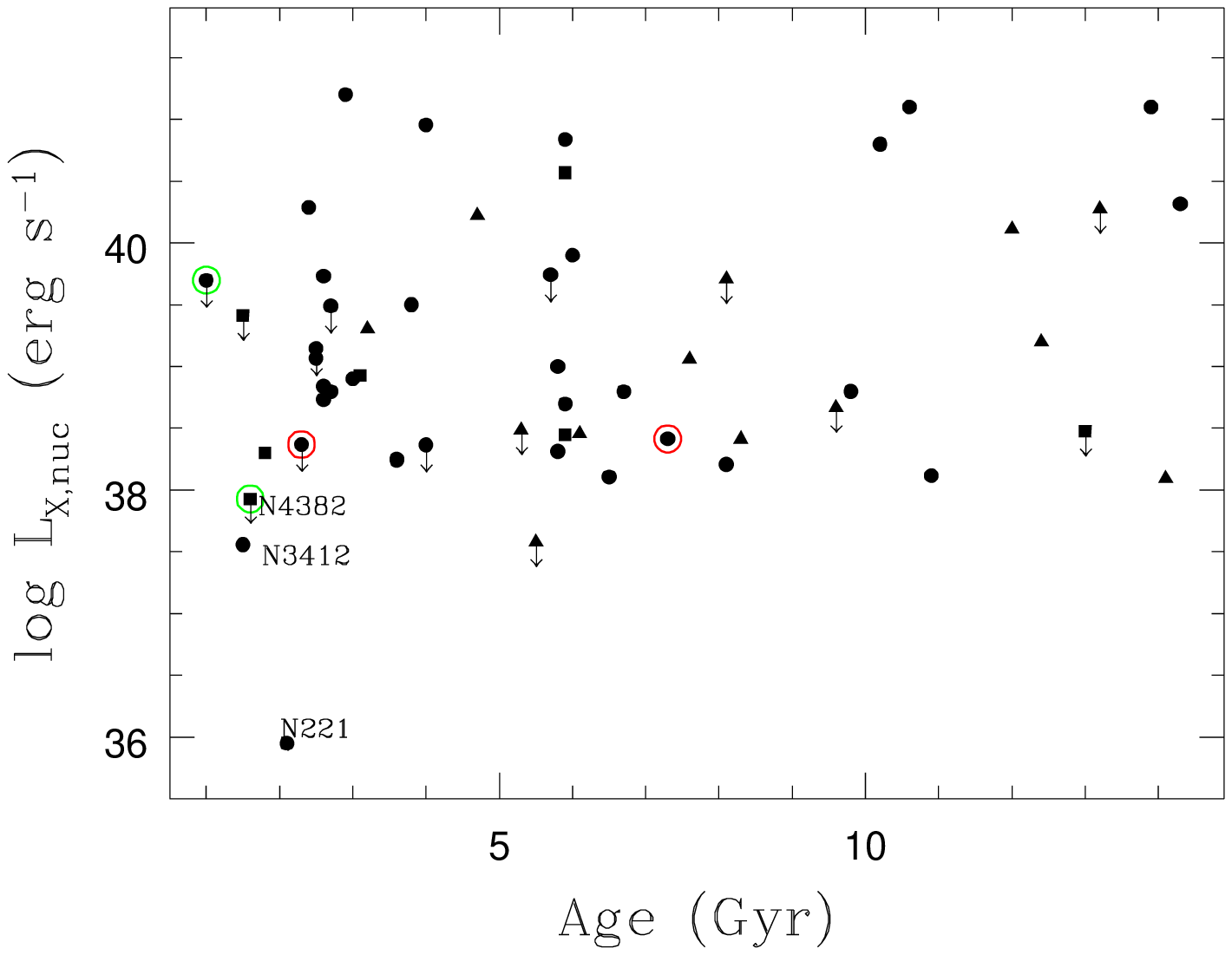}
\vskip -7.5truecm
\hskip 1.truecm
\includegraphics[height=0.75\textheight,width=0.85\textwidth]{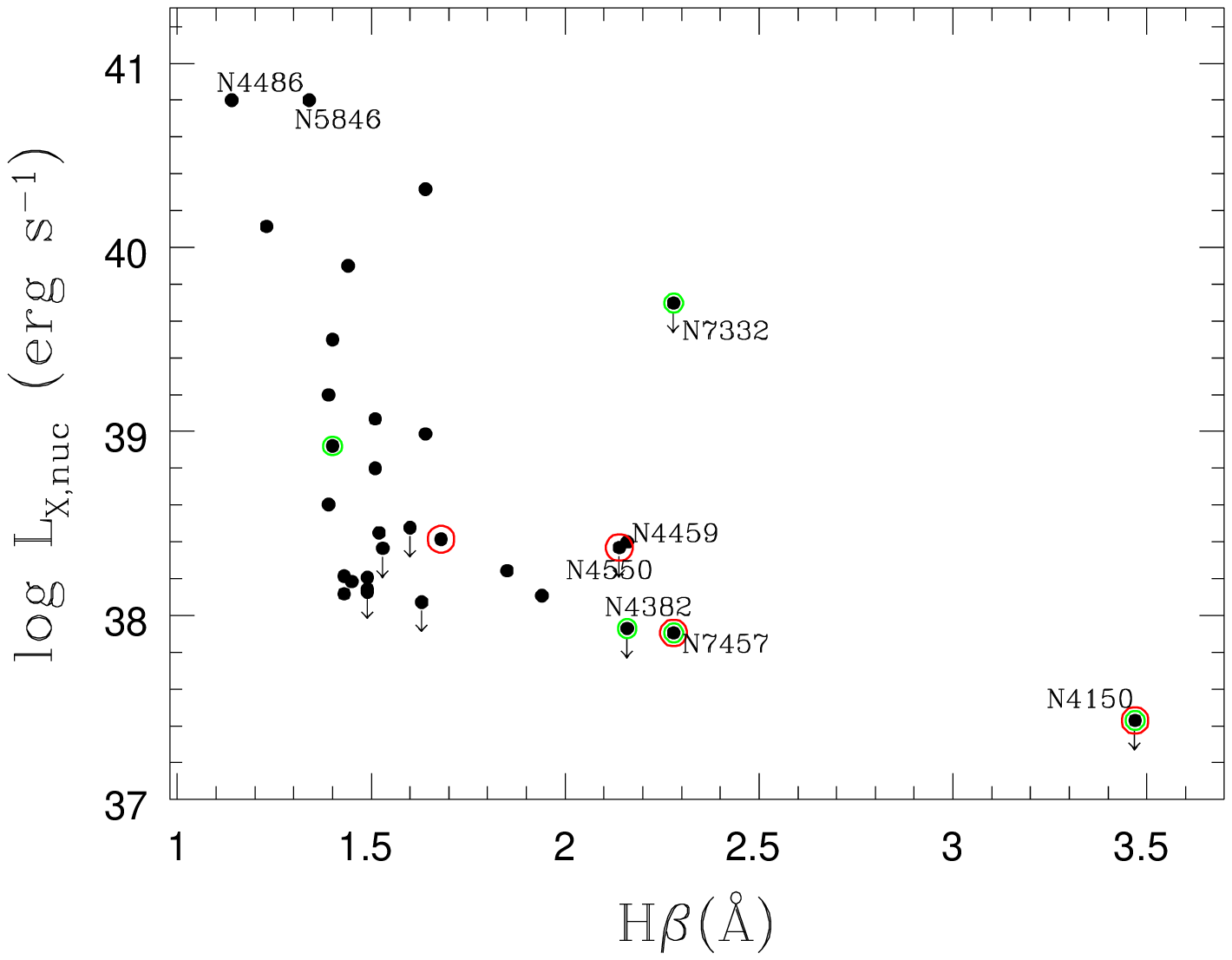}
\vskip -1truecm
\caption{\ln versus stellar population age indicators derived for a
central aperture of radius $R_e/8$: the age of stellar population
synthesis models reproducing observed spectral indeces, in the upper
panel (circles indicate estimates by Denicol\'o et al. 2005, squares
by Terlevich \& Forbes 2002, triangles by Thomas et al. 2005 for an
aperture of $R_e/10$); the H$\beta $ absorption line strength,
indicative of recent starformation, in the lower panel (data
from Kuntschner et al. 2006 for the SAURON sample); see
Sect.~\ref{age} for more details.  In both panels, the red circles
mark the cases where recent starformation has been found from $GALEX$
(Jeong et al. 2009), and the green circles the presence of
kinematically distinct compact cores; in the right panel, the high $H\beta$
line strengths of NGC4150, NGC4382, NGC7332 and NGC7457 are due to these
compact components residing at their centers 
(McDermid et al. 2006).}
\label{f9}
\end{figure*}

The rate of mass return from stars has a dependence on the age of the
stellar population, roughly as $\propto t^{-1.3}$ after an age of
$\sim 2$ Gyrs (e.g., Ciotti et al. 1991). A population age difference
from galaxy to galaxy then leads to a difference in the stellar mass
return rate, that can be comparable to that estimated for variations
in the inner galactic structure in the previous Sect.~\ref{mrate}. For
example, according to the evolutionary population synthesis models of
Maraston (2005), the mass return rate of a simple stellar population
with a Kroupa or Salpeter Initial Mass Function is larger than at an
age of 11 Gyr by a factor of $\sim 1.5$ at 8 Gyr, $\sim 2.3$ at 6 Gyr
and significantly larger for ages $<3$ Gyrs: by a factor of $\sim 9$
at 2 Gyrs, and a factor of $\sim 20$ at 1 Gyr. Most early type
galaxies of the local universe with $M_V<-19$, as for most the sample
in tab.~\ref{tab1}, have ages between 3 and 14 Gyrs, based on single
population evolutionary models applied to spectral line indices
referring to the central galactic region (i.e., within $R_e/10$ or
$R_e/8$; Thomas et al. 2005, Denicol\`o et al. 2005); therefore age
variations can contribute to the wide variation of \ln shown by
figs.~\ref{f1}--\ref{f3} but they will not account for the bulk of it.
Age measurements are available from the references above for $\sim 40$\% of
the sample considered here; using them, \ln shows no clear trend with
age, but instead a constant range of $\sim 3$ orders of magnitude at
all ages from 3 to 14 Gyr (fig.~\ref{f9}, left panel). At smaller
ages, a couple of the lowest-\lb galaxies in the sample (NGC221 and
NGC3412) with their \ln$<10^{38}$ erg s$^{-1}$ extend the \ln range
further towards lower values. Even though the subsample in
fig.~\ref{f9} (left panel) is small and has a bias towards the more
quiescent galaxies, it suggests that the bulk of the \ln variation in
figs.~\ref{f1}-\ref{f3} is not related to age, consistent with the
estimate above of modest variations (within a factor of a few) in the
stellar mass return rate for ages $>3$ Gyr. Numerical models of the
evolution of the nuclear activity in early type galaxies over their
whole lifetime predict a larger frequency of nuclear outbursts in the
past, determined by the larger mass return rate (Ciotti et al. 2010),
and this should produce an increasing presence of high \ln values
going towards smaller ages. Given that in these models the duty cycle
of the nuclear activity is small ($\sim 5\times 10^{-3}$ of the
past 8.5 Gyr), far larger samples are needed to test this prediction
in a quantitative way.

Following $GALEX$ observations, a small fraction ($\sim 10$\%) of
early type galaxies of the local universe has been discovered to have
undergone a recent ($<1-2$ Gyr old) starformation episode (Donas et
al. 2007).  For very large samples, near UV photometry from $GALEX$ in
conjuction with the SDSS optical photometry has provided colors
indicative of low levels (few percent of the total stellar mass) of
recent starformation (age$<$1 Gyr) in $\gsim 30$\% of early type
galaxies of low/intermediate mass ($\sigma <200$ km s$^{-1}$; e.g.,
Schawinski et al. 2007).  It is interesting then to check whether this
phenomenon is related to accretion and \ln\ba. 
Coming to studies of specific objects,
as the SAURON sample including a representative selection of
early type galaxies of the local universe, a $GALEX$ study
revealed recent starformation in $\approx 15$\% of them, again all
with $\sigma <200$ km s$^{-1}$ (Jeong et al. 2009).  The limit $\sigma <
200$ km s$^{-1}$ corresponds to \lb$< 3\times 10^{10}L_{B,\odot}$ for
the galaxies in fig.~\ref{f1}, therefore this recent starformation
phenomenon may be of interest for only a fraction of them.
 High spatial resolution optical spectroscopy of 28 early type SAURON
 galaxies revealed young centers (age $<2$ Gyr) in 6 cases,
 preferentially galaxies that are of low-mass and fast rotating
 (McDermid et al. 2006; Kuntschner et al. 2006).  When considering the
 indicators of recent starformation for the SAURON sample quoted
 above, there could be a trend for the largest \ln to reside 
in galaxies with a
 uniformly old stellar population, typically the most massive ones,
 and the lowest in younger galaxies or those with a younger central
 component, as the kinematically distinct core in NGC4382 (see the \ln
 vs. the H$\beta$ line strength in fig.~\ref{f9}, right panel, where
 galaxies with recent starformation found from $GALEX$ or with 
a kinematically decoupled compact component are
 also evidenced). Figure~\ref{f9} (right panel) suggests a decrease in
 accretion following a starburst episode, or a galaxy merger for which
 the presence of a kinematically distinct component is considered an
 evidence, but unfortunately these considerations are based on too few
 objects to draw definitive conclusions.

Summarizing, the analysis in this Section reveals that 1) at all ages
from 3 to 14 Gyr \ln covers the same wide range of values, without a
clear trend of \ln with age; 2) the lowest \ln values reside in
galaxies with a younger center, recent starformation, and/or with a
younger stellar component at the center, with age $<2-3$ Gyr.  If any,
"youth" seems to be more connected with a lower \ln\ba, but the
following remark is in order.  Most of the cases in 2) (that is
NGC221, NGC3412, NGC4150, NGC4550, NGC4459 and NGC7457, labelled in
fig.~\ref{f9}) are also galaxies with low \lb
[log\lb$(L_{B,\odot})\leq 10.1$], and fig.~\ref{f1} shows a trend of
\ln with \lb\ba; therefore it cannot be concluded whether the lower
nuclear emission is more linked to the galaxy size (\lb\ba) or to the
consequences of recent starformation. The latter may plausibly have a
role in causing a low \ln\ba: during the major galaxy formation
process the feedback from the central MBH is believed to end the
starformation epoch (e.g., Croton et al. 2006), but in the subsequent
evolution recurrent starformation is predicted to take place at the
galactic center, connected with the recurrent nuclear outbursts; soon
after these bursts, the combined heating effects of the starburst and
the central MBH drive the available gas out in a wind, ending abruptly
starformation and accretion as well (Ciotti et al. 2010).  It is
finally interesting that among the cases in 2) there is also the
bright galaxy NGC4382 [log\lb$(L_{B,\odot})\sim 10.8$], likely a
post-merger galaxy due to its kinematically decoupled
component. Post-merger early type galaxies undergo a "rejuvenation" of
their stellar population (e.g., Thomas et al. 2005), as also proved by
X-ray binaries studies (Kim \& Fabbiano 2010), and are typically hot
gas poor (Fabbiano \& Schweizer 1995, Brassington et al. 2007). In
this case, then, heating by starformation or the dynamical effect of the
interaction between galaxies on the hot gas may have emptied the
galaxies and starved the nuclear activity.

\subsection{What determines \ln ?}\label{mdot}

\begin{figure*}
\vskip -10truecm
\includegraphics[height=1.1\textheight,width=1.5\textwidth]{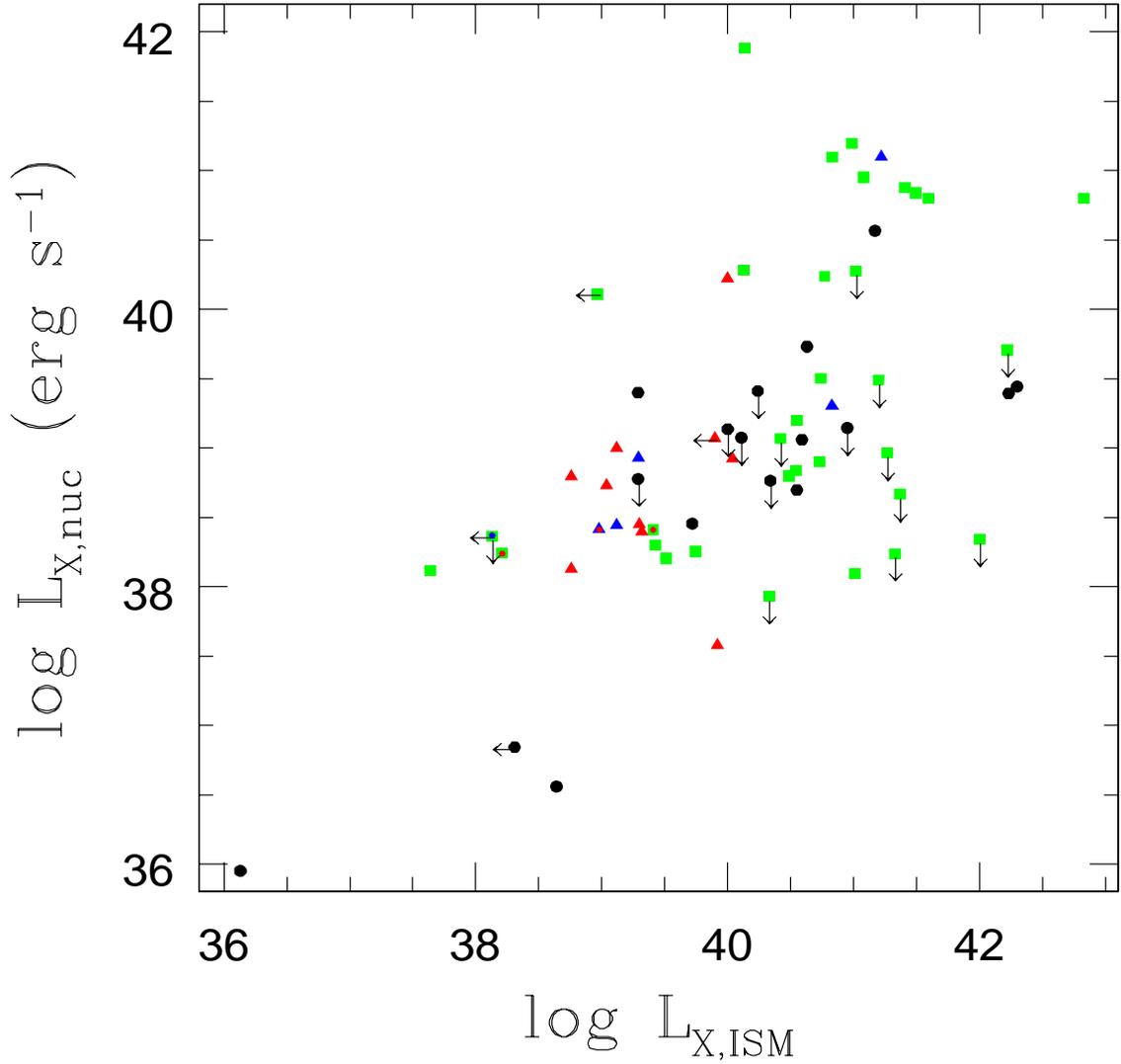}
\vskip -1truecm
\caption{The nuclear luminosity \ln from tab.~\ref{tab1} against the 
X-ray luminosity of the hot ISM estimated from $Chandra$ observations,
from Nagino \& Matsushita (2009), Memola et al. (2009), 
Trinchieri et al. (2008), Diehl \& Statler (2008), Jeltema et al. (2008),
Kim et al. (2008),
Pellegrini et al. (2007a), 
David et al. (2006), Fukazawa et al. (2006), Finoguenov et al. (2006),
Rampazzo et al. (2006),
Kraft et al. (2003), 
Yamasaki et al. (2002).
See Sect.~\ref{mdot}.}
\label{f8}
\end{figure*}

The previous Sects.~\ref{mrate} and~\ref{age} showed that a
variation in the inner stellar structure and in central age, both
strictly related to the stellar mass input rate close to the MBH, can
account for a variation of \md$\,$ of a factor of $\lsim 10$ each;
therefore they can contribute to the wide \ln variation at fixed \lb
or \mb but cannot entirely explain it.  We investigated then whether
\ln has a dependency on the total hot gas emission on the galactic
scale, $L_{X,ISM}$; for galaxies hosting a large steady inflow, the
latter is proportional to the accreting mass flux (e.g., Fabian 2003).
For this purpose we collected the available measurements of the X-ray
emission from hot gas for the galaxies in tab.~\ref{tab1}, based on
$Chandra$ observations, that allow to remove best the contribution of
the nucleus and of X-ray binaries (contaminating instead the
$L_{X,tot}$ values considered in Sect.~\ref{mrate}, from $ROSAT $
data).  These gas luminosities $L_{X,ISM}$, for 65 of the objects in
tab~\ref{tab1}, were converted to the same (0.3--2) keV band using the
spectral shape used to derive them, and to the distances in
tab.~\ref{tab1}, and then compared with \ln
in fig.~\ref{f8} (references are given in its caption).
Even though less objects are present 
than in figs.~\ref{f1} and~\ref{f2}, fig.~\ref{f8} shows that \ln is
detected both in gas poor and in gas rich galaxies, and on average
increases with $L_{X,ISM}$, but with a wide variation of \ln of $2- 4$
orders of magnitude for log$L_{X,ISM}($erg s$^{-1})\gsim 39$.

Evidently, the level of radiative output from accretion is not tightly
related to fuel availability within a steady scenario, either on the
circumnuclear or on the galactic scale.  Alternatively, then,
accretion could be intermittent due to feedback from the central MBH,
that occasionally heats the surrounding gas and lowers \md$\,$ until
accretion is stopped; later on cooling resumes and \md$\,$ can
increase again, even above values predicted for a steady inflow, since
gas that has been pushed and accumulated at large radii can fall back
towards the center again (e.g., Di Matteo et al. 2003, Ciotti,
Ostriker \& Proga 2010). In support for this, the hot gas morphology
of a large fraction of high $L_{X,ISM}$ galaxies shows cavities,
shells and shocks, often associated with radio jets and lobes, that
can be attributed to past nuclear outbursts, taking place every few
$10^6-10^8$ yrs (e.g., Forman et al. 2005, Allen et al. 2006, Diehl \&
Statler 2008, Baldi et al. 2009).  The large spread in \ln could then
be produced by the activity cycle, where \ln is regulated by the joint
actions of feedback and fuel availability.

In addition to heating from feedback, the \md$\,$ reaching the MBH
could be regulated by another phenomenon, for which there are both
theoretical and observational evidences.  In RIAFs, only a small
fraction of the gas supplied may actually fall on to the MBH, and the
binding energy it releases may be transported radially outward and
drive away the remainder in a wind (Blandford \& Begelman 1999). The
loss of a large fraction of the accretion flow is also possible due to
angular momentum (Proga \& Begelman 2003a) or a weak magnetic field
(Proga \& Begelman 2003b, Hawley \& Balbus 2002). On the observational
side, detailed studies of ellipticals based on $HST$ and $Chandra$ to
estimate the accretion radius $r_a$, the mass accretion rate within
$r_a$ and \ln\ba, concluded that for many of them \ln is so low that
most of the available gas within $r_a$ cannot reach the MBH, even when
allowing for a low radiative efficiency (Fabbiano et al. 2004;
Pellegrini 2005b; Soria et al. 2006b; Ho 2009). This "fuel
overabundance" problem has been best investigated for Sgr A$^*$, where
the X-ray and radio results imply outflows or convection close to the
MBH (Baganoff et al. 2003). The possibility of a large mass loss for
the accretion flow entering $r_a$ before it reaches the MBH could be
another factor accounting for the wide variation in \ln\ba, if taking
place in different proportions at fixed \mb\ba.  It would also help
explain the mild (if any) sensitivity of \ln to different
circumnuclear fuel productions discussed in Sects.~\ref{obs}
and~\ref{mrate}, since the possibility of a large accretion flow
depletion overcomes the role of cuspiness in determining \md. A
variation in the fraction of the accreting mass at $r_a$ that
effectively reaches the MBH could have already been observed: together
with the "fuel overabundance" cases quoted above, in some hot gas rich
galaxies with cavities the mass lost by the accretion flow on its way
to the MBH cannot be large: from the energy input by feedback to their
hot coronae, a high efficiency was found ($\sim 1/5$) for the
conversion into jet power of the accretion power $L_{acc}\sim 0.1
$\md$c^2$, where \md$\,$ is the gas mass entering $r_a$ assuming a
Bondi rate (Allen et al. 2006).

Finally, in figs.~\ref{f1}--\ref{f3} there seems to be no gap in \ln
and \ln\ba/\ledd between the brightest objects, that are a few
Seyferts, and the other more radiatively quiescent objects.  If the
low luminosity Seyferts in this sample share the same accretion
mechanism of the brighter Seyferts that are classified as classical
AGNs, i.e., they are powered by a standard thin accretion disk plus
hot corona (Maoz 2007; Panessa et al. 2007), then there must be a smooth
transition from radiatively efficient to inefficient accretion. This
may indicate a gradual changeover from a pure RIAF to an inner RIAF
plus an outer thin disk, to a thin disk plus corona system (as
suggested, e.g., by Ho 2008 for the Palomar sample).  A continuous
distribution in nuclear X-ray emission from $10^{38}$ to $10^{42}$ erg
s$^{-1}$ was also found for the sample of nearby galaxies, of type Sa
or later (Zhang et al. 2009).

\subsection{The radio and X-ray properties of nearby nuclei}\label{difxr}

\begin{figure*}
\hskip 0.8truecm
\includegraphics[height=1\textheight,width=1\textwidth]{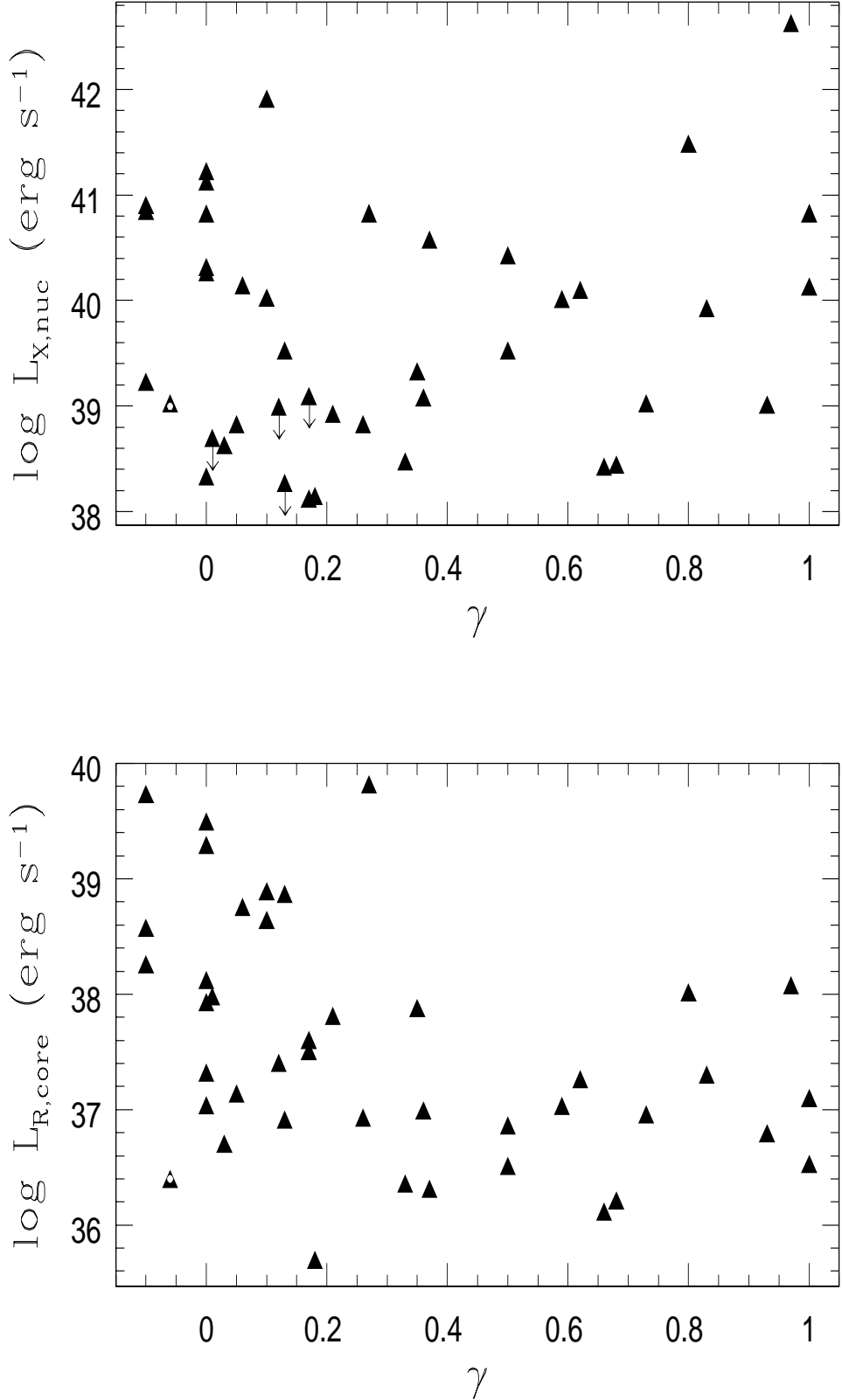}
\caption{The relationship between the inner light profile shape $\gamma$ and
\ln (upper panel), or the core radio luminosities at 5 GHz 
(lower panel),
for the Capetti \& Balmaverde (2005) sample.
The L-shaped trend with $\gamma$ is apparent for the radio luminosity,
but is absent for \ln\ba.
The white dot inside the symbol
marks NGC3945 that becomes a cusp galaxy according to $\gamma^{\prime}$.
See Sect.~\ref{xr} for more details.}
\label{f6}
\end{figure*}

We discuss here the different behavior of \ln and the core 5 GHz
luminosity $L_{R,core}$ with respect to $\gamma$ (Sect.~\ref{xr}).
The trend in the $L_{R,core}$ versus $\gamma$ plot (fig.~\ref{f6}) is
similar to the L-shape in the plot of the total soft X-ray emission
$L_{X,tot}$ versus $\gamma$ (Pellegrini 1999, 2005a): core galaxies
span a large range of soft X-ray emission, from the lowest to the
highest values observed, while cusp galaxies are confined below
$L_{X,tot}\sim 10^{41}$ erg s$^{-1}$.  The relationship was explained
with the core profile being characteristic of the more massive
galaxies, often centrally located in groups and clusters, and then
with the most favorable conditions to retain the hot gas.

The similar trends of $L_{R,core}$ and $L_{X,tot}$ with respect to
$\gamma$ are not unexpected, given that these two quantities correlate
each other (Fabbiano et al. 1989).  The L-shape that they show with
respect to cuspiness could then be produced in both cases by the hot
gas, determining it directly $L_{X,tot}$ and being also essential for
jet confinement and propagation (e.g., Worrall 2002, Kaiser
2009). In fact many of the core galaxies in fig.~\ref{f6} have a
radio morphology with well developed jets and lobes, or an extended
radio structure indicative of a collimated outflow (Capetti \&
Balmaverde 2006), and the VLA radio luminosities include a
contribution from the jet components. When $L_{R,core}$ and
$L_{X,ISM}$ are above their respective thresholds, then, galaxies have
the most dense coronae and the related action of confinement is more
efficient. An additional link between the hot gas and the radio
emission could be due to a higher accretion rate produced by a higher
gas cooling rate $L_{X,ISM}$ (as already suggested previously,
Fabbiano et al. 1989; Mittal et al. 2009). Likely both actions are at
work, that is a higher hot gas content provides both a more effective
confinement and a higher accretion rate.  The latter should of course
produce also a higher \ln\ba, and indeed the investigation performed
here (fig.~\ref{f8}) shows a correlation between the nuclear
luminosity \ln and $L_{X,ISM}$. The L-shape, though, is present
only with respect to $L_{R,core}$, not \ln (fig.~\ref{f6}), probably
because the action of confinement is more important for 
for $L_{R,core}$ than  \ln\ba. However, a more tight
relationship between the (Bondi) accretion rate and the outcome of
accretion has been found considering the jet power rather than the
nuclear luminosities [Sect.~\ref{mdot}; see also the lack of
significant correlation between \ln and the Bondi rate, Pellegrini
(2005b), Merloni \& Heinz (2007)] since in nearby low luminosity or
almost quiescent nuclei the output of accretion is dominated by the
kinetic rather than the radiative power [K\"ording et al. 2006, Allen
et al. 2006, Merloni \& Heinz (2007)].  An implicit consequence of the
different trend of the radio and nuclear X-ray luminosities in
fig.~\ref{f6} is that the $L_{R,core}/$\ln ratio should have the
largest values among core galaxies, and keep below a threshold for
power law galaxies. In fact, power law galaxies tend to be radio quiet
and core galaxies radio loud, when using the radio-loudness parameter
$R_X$=L(5 GHz)/L(2--10 keV) of Terashima \& Wilson (2003) [see also
Capetti \& Balmaverde (2006); Panessa et al. (2007)].

Finally, it seems (fig.~\ref{f6}) that only core galaxies can reach
the highest $L_R$ and possess a conspicuous radio activity cycle.  An
activity cycle in cusp galaxies may take place, but with a variation
of the radio emission that keeps within a smaller range, perhaps
because of a rapid jet failure due to the lack of dense confining
gas. For example in the hot gas poor NGC821 a mini-jet in the nuclear
region has been possibly discovered in the radio (Pellegrini et
al. 2007b).  Another possibility is given by the fact that cusp
galaxies are on average less massive, and the duty cycle likely
increases with galaxy mass, because single outbursts have greater
impact in less massive systems (Ciotti et al. 2010), which then are
"on" for a shorter time. Evidence for this is being found
from a large sample of hot gas coronae studied with $Chandra$, where
the duty cycle seem to increase from $<10$\% in the less luminous
galaxies with less massive hot gas halos to $>50$\% in the most
luminous ones (Nulsen et al. 2009).

\section{Conclusions}\label{concl}

In this work measurements or upper limits have been collected for the
hard X-ray emission at the nucleus of 112 early type galaxies (E and
S0) of the local universe (within 67 Mpc). \ln derives from $Chandra$
data for 94\% of the nuclei, and the sample includes all the
available measurements of \ln (29 cases) for galaxies with the above
characteristics and a direct estimate of the \mb (38 objects). Using
also \mb from the \mb$-\sigma$ relation, and the inner stellar profile
(slope $\gamma$ and break radius $r_b$) measured with $HST$ for 81
galaxies, the relationships between \ln and \mb\ba, the central
stellar structure, the central age, the radio and the soft X-rays (hot gas)
luminosities have been investigated, with the following results:

$\bullet$ \ln increases on average with \lb and \mb\ba, with a wide
variation of \ln and \ln\ba/\ledd\ba, up to 4 orders of magnitude, at
any fixed galactic \lb$> 6\times 10^{9}L_{B,\odot}$ or
\mb$>10^7M_{\odot}$. Cusp and core galaxies both cover the whole large
range of \ln\ba, without a clear trend with $\gamma$ or $r_b$.  Most
nuclei have an Eddington ratio $-9<$log(\ln\ba/\ledd\ba)$<-5$, the
brightest ones (with log(\ln\ba/\ledd\ba)$\sim -4$) are four Seyfert
nuclei. The  \ln\ba/\ledd is highest at the lowest \mb\ba, and
shows an increase of its highest values with increasing "cuspiness" ($\gamma$).

$\bullet$ Accretion in these MBHs may have entered the
radiatively inefficient regime, where \ln scales as the mass accretion
rate \md$^2$ at fixed \mb\ba; therefore reasons for \md$\,$ variations
explaining the wide range of observed \ln at fixed \lb (or \mb\ba) are
searched for.  In a scenario where accretion is (quasi) steady,
\md$\,$ could vary because of a different inner galactic structure,
that determines the amount of mass shed by stars in the circumnuclear
region and then available for accretion. The lack of a clear trend of
\ln with $\gamma$ and $r_b$, and the weak evidence for an increase of
\ln\ba/\ledd with cuspiness, indicate that differences in the inner
galactic structure do not have the dominant effect on \ln\ba. In fact
the circumnuclear fuel production rate is estimated to vary more than
a factor of a few only when the inflowing region is small (of the
order of the accretion radius) and for large differences in $r_b$; in
agreement with this, for galaxies with presumably a small inflowing region,
a trend for \ln to decrease for increasing $r_b$ seems to be
present.

$\bullet$ The stellar mass return rate depends also on the age of the
stellar population, but \ln covers the same wide range of values at
all central ages from 3 to 14 Gyr; "youth", indicated by a younger
center, recent starformation, and/or a younger stellar component at
the center (of ages $<2-3$ Gyr), seems to be more connected with a
lower \ln\ba. An explanation could be that the heating by
starformation or the dynamical effect of the merging process on the
hot gas drive the available gas out in a wind, ending abruptly the
starformation and accretion.  However, this finding remains an
indication, because of the trend of \ln with \lb and the small number
of galaxies with information about recent starformation episodes.

$\bullet$ \ln on average increases with the total hot gas emission
$L_{X,ISM}$, but the relation is not tight: it shows a large variation
of 2--3 orders of magnitude both at the lowest and highest gas
contents.  Nuclear emission at the highest detected levels for this
sample (\ln$\sim 10^{40}-10^{41}$ erg s$^{-1}$) is present even when
the gas content is low, and \ln at the lowest levels ($\sim
10^{38}-10^{39}$ erg s$^{-1}$) can be found at gas luminosities differing by 3
orders of magnitude.

$\bullet $ The mild sensitivity of \ln to the circumnuclear and
global hot gas contents, and its large variation, finds two possible
explanations, both of which go in the sense of overcoming the
importance of fuel availability : 1) the gas is heated due to feedback
from the MBH, which could be a phenomenon not limited to high
$L_{X,ISM}$ galaxies, or 2) only a small fraction of the mass entering
the accretion radius actually reaches the MBH, as in RIAF models with
winds/outflows, and this fraction can vary largely at fixed \mb\ba.

$\bullet $ A sub-sample of galaxies shows an already known trend by
which cusp galaxies are confined below a threshold in the VLA 5 GHz
central luminosity $L_{R,core}$, while core galaxies span a large
range of $L_{R,core}$; this subsample does not show a similar trend of
\ln with the central light profile.  The $L_{R,core}$ behavior is
instead similar to that of the total soft X-ray emission with respect
to $\gamma$; the hot gas could then be responsible both for the soft
X-ray emission and the jet confinement and propagation. While
core galaxies can possess a conspicuous radio activity cycle, in cusp
galaxies the variation of the radio emission keeps within a smaller
range, because of a rapid jet failure due to the lack of a dense
confining medium, or a smaller duty cycle, being these galaxies 
on average less massive systems.

\acknowledgments

The referee is thanked for useful comments.
I acknowledge use of the NASA Extragalactic database (NED), operated by the
Jet Propulsion Laboratory, California Institute of Technology, and of the
Hyperleda database (http://leda.univ-lyon1.fr/).

\begin{table*}
\caption[] { The nuclear properties of the sample of early type galaxies.}\label{tab1}
\begin{tabular}{ l  r r r r r r r r  }
\noalign{\smallskip}
\hline
\noalign{\smallskip}
 Name   & $B_T^0$   & d & Ref&$\gamma$ & log \ln  & Ref & 
log $M_{BH}$ & Ref  \\
        & (mag)     & (Mpc) &    &           & (erg s$^{-1})$    &     & ($M_{\odot}$) &  \\
 (1)  &   (2)          & (3)& (4)   &  (5)         &   (6)                 & (7)& (8)        & (9) \\
\noalign{\smallskip}
\hline
\noalign{\smallskip}
NGC221  &    8.22   &    0.81  &  a  & ..... & 35.95 & 1 & 6.47  &  2 \\
NGC404  &    10.94  &     3.3  &  a  & ..... & 36.84 & 2 & 5.39  &  1 \\
NGC474  &    12.20  &    29.3  &  b  &  0.37 & 38.41 & 3 & 7.88  &  1 \\
NGC507  &    12.23  &    63.8  &  b  &  0.00 &$<$39.71&4 & 8.96  &  1 \\
NGC524   &   11.01  &    24.0  &  a  &  0.03 & 38.60 & 3 &  8.63  & 1  \\
NGC720  &    11.05  &    27.7  &  a  &  0.06 & 38.90 & 5 & 8.56  &  1 \\
NGC821  &    11.25  &    24.1  &  a  &  0.10 &$<$38.37&6 & 7.60  &  2 \\
NGC1023  &    9.29  &    11.4  &  a  &  0.74 & 38.13 & 7 & 7.64  &  2 \\
NGC1052  &   11.32  &    19.4  &  a  &  0.18 & 41.20 & 3 & 8.29  &  1 \\
NGC1316  &    9.29  &    21.5  &  a  &  0.35 & 39.30 & 8 & 8.44  &  1 \\
NGC1331  &   14.15  &    24.2  &  a  & ..... & 38.13 & 9 & 6.09  &  1 \\
NGC1332  &   11.03  &    24.2  &  a  & ..... &$<$38.77&9 & 9.04  &  1 \\
NGC1380  &   10.80  &    17.6  &  a  & 1.0$^a$& 40.10 &10 & 8.40  &  1 \\
NGC1395  &   10.48  &    24.1  &  a  & ..... & 39.06 & 11& 8.59  &  1 \\
NGC1399  &   10.35  &    20.0  &  a  &  0.12 &$<$38.96&12 & 8.68  &  2 \\
NGC1404  &   10.81  &    21.0  &  a  & ..... & 40.57 & 10 & 8.50  &  1 \\
NGC1407  &   10.38  &    28.8  &  a  &.....  &$<$39.14&14 & 8.76  &  1 \\
NGC1549  &   10.61  &    19.7  &  a  &.....  & 38.46 & 13 & 8.25  &  1 \\
NGC1553  &   10.20  &    18.5  &  a  &  0.74 & 40.22 & 2 & 8.02  &  1 \\
NGC1600  &   11.68  &    57.4  &  b  & -0.03 &$<$39.49&15 & 9.11  &  1 \\
NGC1700  &   11.78  &    40.6  &  b  & -0.10 & 38.84 & 13 & 8.54  &  1 \\
NGC2110  &   11.85  &    31.3  &  c  & ..... & 42.47 & 16 & 8.65 & 1 \\
NGC2300  &   11.73  &    30.4  &  b  &  0.07 & 40.96 & 10 & 8.69  &  1 \\
NGC2434  &   11.25  &    21.6  &  a  &  0.75 &$<$37.58& 10 & 8.10  &  1 \\
NGC2865  &   12.05  &    37.8  &  a  &.....  & 39.40 & 17 & 7.97  &  1 \\
NGC2974  &   11.61  &    21.5  &  a  &  0.62 & 40.32 & 10 & 8.53  &  1 \\
NGC3115  &    9.87  &     9.7  &  a  &  0.52 & 38.73 & 7 & 8.96  &  2 \\
NGC3125  &   13.01  &    13.4  &  c  &.....  & 37.60 & 18 & 5.91  &  1 \\
NGC3193  &   11.86  &    34.0  &  a  &  0.01 &$<$39.74& 10 & 8.17  &  1 \\
NGC3226  &   12.22  &    23.6  & a  & 1.0$^a$  & 40.80 & 3 & 8.17  &  1 \\
NGC3245  &   11.53  &    20.9  & a  & 0.73$^a$ & 39.00 & 3 & 8.32  &  2 \\
NGC3309  &   12.15  &    56.5  &  c  &.....  &$<$39.14& 19 & 8.67  &  1 \\
NGC3311  &   12.40  &    52.7  &  c  &.....  &$<$39.08& 19 & 8.10  &  1 \\
NGC3377  &   10.98  &    11.2  &  a  &  0.03 & 38.24  & 20 & 8.02  &  2 \\
NGC3379  &   10.11  &    10.6  &  a  &  0.18 & 38.12  & 21 & 8.04  &  2 \\
NGC3384  &   10.76  &    11.6  &  a  &  0.71 & 38.11  & 7  & 7.24  &  2 \\
NGC3412  &   11.30  &    11.3  &  a  &  .....& 37.56  & 7  & 7.06  &  1  \\
NGC3414  &   11.93  &    25.2  &  a  &  0.83 & 39.90  & 3  & 8.52  &  1 \\
NGC3516  &   12.32  &    35.0  &  c  &0.97$^a$& 42.60 & 16 & 7.82  &  1 \\
NGC3557  &   10.92  &    45.7  &  a  &0.00$^a$& 40.24 & 22 & 8.77  &  1 \\
NGC3585  &   10.52  &    20.0  &  a  &  0.31 & 38.93  & 23 & 8.51  &  2 \\
NGC3607  &   10.83  &    22.8  &  a  &  0.26 & 38.80  & 3  & 8.14  &  2 \\
NGC3608  &   11.46  &    22.9  &  a  &  0.09 & 38.21  & 3  & 8.32  &  2 \\
NGC3923  &   10.39  &    22.9  &  a  &.....  & 39.73& 10 &8.65  &  1 \\
NGC3945  &   11.56  &    19.9  &  b  & -0.06 & 39.00  & 3   & 8.00  & 1 \\  
NGC3998  &   11.32  &    14.1  &  a  &0.80$^a$& 41.46 & 24 & 8.36  &  2 \\
NGC4026  &   11.59  &    13.6  &  a  &0.68$^a$& 38.41 & 18 & 8.26  &  2 \\
NGC4036  &   11.41  &  23.3  & c     &0.36$^a$& 39.05 & 3 & 8.06  &  1 \\
NGC4111  &   11.60  &  15.0  & a     &0.50$^a$& 40.40 & 3 & 7.71  &  1 \\
NGC4125  &   10.53  &    23.9  &  a  &.....   & 38.70 & 3 & 8.45  &  1 \\
NGC4143  &   11.75  &    15.9  &  a  &  0.59  & 39.98 & 2 & 8.35  &  1 \\
NGC4150  &   12.41  &    13.7  &  a  &  0.58  &$<$37.43&7 & 6.76  &  1 \\
NGC4168  &   11.92  &    37.3  &  b  &  0.17  &$<$39.07&22& 8.09  &  1 \\
NGC4203  &   11.66  &    15.1  &  a  &0.62$^a$ & 40.08 & 7 & 7.87  &  1 \\
NGC4261  &   11.24  &    31.6  &  a  &  0.00 & 41.10 & 3 & 8.72  &  2 \\
NGC4278  &   10.91  &    16.1  &  a  &  0.06 & 40.11 & 25 & 8.53  &  1 \\
NGC4291  &   12.20  &    26.2  &  a  &  0.01 & 40.57 & 10 & 8.53  &  2 \\
NGC4342  &   13.36  &    11.6  &  c  &.....  & 38.70 & 18 & 8.37  &  2 \\
NGC4365  &   10.39  &    20.4  &  a  &  0.07 & 38.25 & 26 & 8.65  &  1 \\
NGC4374  &    9.89  &    18.4  &  a  &  0.13 & 39.50 & 3  & 9.20 &  2 \\
NGC4382  &    9.88  &    18.4  &  a  &  0.00 &$<$37.93&26 & 8.04  &  1 \\
NGC4387  &   12.82  &    21.4  &  a  &  0.10 &$<$38.48&26 & 7.12  &  1 \\
NGC4406  &    9.71  &    17.1  &  a  & -0.04 &$<$38.34&26 & 8.51  &  1 \\
\hline
\end{tabular} 
\end{table*}

\begin{table*}
\begin{tabular}{ l  r r r r r r r r  }
\noalign{\smallskip}
\hline
\noalign{\smallskip}
 Name   & $B_T^0$   & d & Ref&$\gamma$ & log \ln & Ref & 
log $M_{BH}$ & Ref  \\
        & (mag)     & (Mpc) &    &           & (erg s$^{-1})$    &     & ($M_{\odot}$) &  \\
 (1)  &   (2)          & (3)& (4)   &  (5)         &   (6)                 & (7)& (8)        & (9) \\
\noalign{\smallskip}
\hline
\noalign{\smallskip}
NGC4417  &   11.96  &    16.7  &  a  &  0.71 &$<$38.26&26 & 7.55  &  1 \\
NGC4435  &   11.37  &    17.4  &  a  & 0.33$^a$ & 38.45 & 26 & 7.81  &  1 \\
NGC4458  &   12.77  &    17.2  &  a  &  0.16 &$<$38.07 & 26&7.08  &  1 \\
NGC4459  &   11.25  &    16.1  &  a  & 0.66$^a$ & 38.40 & 3  & 7.85  &  2 \\
NGC4464  &   13.45  &    16.5  &  a  & ..... & 38.39 & 26 & 7.45  &  1 \\
NGC4467  &   14.79  &    17.2  &  a  & ..... &$<$38.45& 26 & 6.42  &  1 \\
NGC4472  &    9.16  &    16.3  &  a  &  0.01 &$<$38.67& 12 & 8.89  &  1 \\
NGC4473  &   10.91  &    15.7  &  a  & -0.07 &$<$38.14& 7  & 8.08  &  2 \\
NGC4478  &   12.04  &    18.1  &  a  & -0.10 &$<$38.49& 26 & 7.59  &  1 \\
NGC4486  &    9.47  &    16.1  &  a  &  0.27 & 40.80  & 3  & 9.53  &  2 \\
NGC4486B  &  14.18  &    16.9  &  a  & ..... & 37.91  & 20 & 7.95  &  1 \\
NGC4494  &   10.57  &    17.1  &  a  &  0.52 & 38.80  & 3  & 7.74  &  1 \\
NGC4550  &   12.31  &    15.9  &  a  &.....  &$<$38.37& 7  & 6.98  &  1 \\
NGC4552  &   10.49  &    15.4  &  a  & -0.10 &  39.20 & 3  & 8.63  &  1 \\
NGC4564  &   11.81  &    15.0  &  a  &  0.80 &  38.45 & 20 & 7.78  &  2 \\
NGC4570  &   11.62  &    17.9  &  a  & ..... &  38.18 & 26 & 8.12 &  1 \\
NGC4578  &   12.24  &    18.5  &  a  & ..... &$<38.36$& 26 & 7.35 &  1  \\
NGC4589  &   11.55  &    22.0  &  a  &  0.21 &  38.90 & 3  & 8.43  &  1 \\
NGC4612  &   11.94  &    17.2  &  a  & ..... &  38.08 & 26 & 6.24 &  1  \\
NGC4621  &   10.52  &    18.3  &  a  &  0.75 &  38.92 & 7  & 8.44  &  1 \\
NGC4636  &   10.29  &    14.7  &  a  &  0.13 &$<$38.24& 12 & 8.25  &  1 \\
NGC4649  &    9.70  &    16.8  &  a  &  0.17 &  38.09 & 7  & 9.33  &  2 \\
NGC4660  &   11.90  &    12.8  &  a  &  0.91 &  38.22 & 26 & 8.13  &  1 \\
NGC4696  &   11.13  &    35.5  &  a  &  0.10 &  40.00 & 3  & 8.64  &  1 \\
NGC4697  &   10.10  &    11.8  &  a  &  0.22 &  38.41 & 20 & 8.28  &  2 \\
NGC4754  &   11.35  &    16.8  &  a  &.....  &  38.31 & 26 & 8.09  &  1 \\
NGC4759  &   13.75  &    50.7  &  c  & ..... &$<38.78$& 27 & 8.49  & 1  \\
NGC5018  &   11.23  &    39.4  &  c  &.....  &$<39.41$& 28 & 8.32  &  1 \\
NGC5044  &   11.24  &    31.2  &  a  &.....  &  39.44 & 29 & 8.54  &  1 \\
NGC5102  &   10.01  &    4.0   &  a  & ..... &  36.56 & 18 & 7.58  &  1 \\
NGC5128  &    7.28  &    4.2   &  a &0.10$^a$&  41.88 & 30 & 8.46  &  2 \\
NGC5273  &   12.47  &    16.5  &  a &0.37$^a$&  40.55 & 31 & 6.60  &  1 \\  
NGC5283  &  14.15   &    48.7  &  c &..... &    41.91 & 32 & 7.71 &  1  \\
NGC5322  &   10.96  &    31.2  &  a &0.00$^a$&  40.29 & 10 & 8.49  &  1 \\
NGC5419  &   11.53  &    62.6  &  b  & -0.10 &  40.82 & 22 & 9.20  &  1 \\
NGC5813  &   11.24  &    32.2  &  a  &  0.05 &  38.80 & 3  &  8.52  &  1 \\
NGC5838  &   11.53  &    22.2  &  b  &  0.93 &  38.99 & 3  & 8.72  &  1 \\
NGC5845  &   13.19  &    25.9  &  a  &  0.51 &  39.07 & 20 & 8.42  &  2 \\
NGC5846  &   10.82  &    24.9  &  a  &0.00$^a$& 40.80 & 3  & 8.54  &  1 \\
NGC5866  &   10.66  &    15.3  &  a  &0.00$^a$ & 38.30 & 3  & 7.84  &  1 \\
NGC6482  &   11.73  &    58.6  &  c  & .....  &  39.39 & 3  & 8.97  &  1 \\
NGC7052  &   13.30  &    67.1  &  b  &   0.16 &$<$40.28& 4  & 8.58  &  2 \\
NGC7332  &   11.79  &    23.0  &  a  &   0.62 &$<$39.70& 10 & 7.42  &  1 \\
NGC7457  &   11.63  &    13.2  &  a  & -0.10 &  37.91 & 18  & 6.58  &  2 \\
NGC7619  &   11.70  &    53.0  &  a  & -0.02 &  40.84 & 7 & 9.05  &  1 \\
NGC7626  &   11.81  &    53.0  &  a  &  0.36 &  41.10 & 7 & 8.75  &  1 \\
NGC7743  &   12.07  &    20.7  &  a  &  0.50 &  39.50 & 3 & 6.75  &  1 \\
IC1459  &    10.86  &    29.2  &  a  & -0.10 &  40.87 & 33 & 9.42  &  2 \\
IC4296   &    11.24 &    48.8  &  a & 0.00$^a$ &  41.20 & 34 & 9.10  &  1 \\
\hline
\end{tabular} 

\bigskip
$^a$ $\gamma$ from the modeling with the Nuker law by
Capetti \& Balmaverde (2005), see Sect.~\ref{obs}; NGC1380 and NGC3226
are just defined cusp galaxies, for them $\gamma=1$ is assumed here.

\bigskip

Column (1): galaxy name. Col. (2): total apparent blue magnitude,
corrected for galactic and internal extinction, from
HyperLeda. Col. (3): distance, for $H_0=70$ km s$^{-1}$ Mpc$^{-1}$,
based on sources given in col. (4), taken with this priority: the SBF
method [Tonry et al. (2001), a in col. (4); distances derived from
this work are now consistent with $H_0=70$ km s$^{-1}$ Mpc$^{-1}$
after more recent Cepheid and Hubble flow recalibrations, J. Tonry
2010, private communication; for IC4296 the reference is Mei et 
al. 2000; for a few Virgo galaxies not in Tonry et al. (2001), the results of
Blakeslee et al. (2009) have been used]; Lauer et al. (2007a) [b in col. (4); 
these authors reworked the distances in various sources to be consistent
with $H_0=70$ km s$^{-1}$ Mpc$^{-1}$]; the recession velocity
corrected for Virgo infall, given by HyperLeda [c in col. (4)]. Col. (5): 
inner slope of stellar light profile, from Lauer
et al. (2007a), except for cases marked with the apex $^a$.  Col. (6):
nuclear luminosity in the 2--10 keV band, rescaled for the distance in
col.(3), from the reference in col. (7) (see Sect.~\ref{sample}).
Col. (8): central black hole mass, from the method specified in
col. (9), where 1 = the \mb$-\sigma$ relation of G\"ultekin et
al. (2009) for elliptical galaxies, 2 = a direct mass measurement with
a dynamical modeling (references in G\"ultekin et al. 2009); when necessary,
the masses have been rescaled for the distance in col. (3).

\smallskip

\tablerefs{ col. (7):
1=Ho et al. 2003,
2=Eracleous et al. 2010,
3=Gonz\'alez-Mart\'in et al. 2009,
4=Donato et al. 2004, 
5=Jeltema et al. 2003, 
6=Pellegrini et al. 2007b, 
7=Ho 2009,
8=Rinn et al. 2005, 
9=Humphrey \& Buote 2004, 
10=Liu \& Bregman 2005, or Liu 2008 (arXiv:0811.0804) for NGC1404, NGC2434 and NGC3923,
11=Colbert et al. (2004),
12=Loewenstein et al. 2001, 
13=Diehl \& Statler 2008, 
14=Zhang et al. 2007, 
15=Sivakoff et al. 2004, 
16=Winter et al. 2009,
17=Sansom et al. 2006,
18=Zhang et al. 2009, 
19=Yamasaki et al. 2002, 
20=Soria et al. 2006a, 
21=Brassington et al. 2008, 
22=Balmaverde \& Capetti 2006,
23=O'Sullivan \& Ponman 2004,
24=Pellegrini et al. 2000,
25=Brassington et al. 2009,
26=Gallo et al. 2010, 
27= Morita et al. 2006,
28=Ghosh et al. 2005, 
29=David et al. 2009,
30=Evans et al. 2004, 
31=Capetti \& Balmaverde 2006,
32=Ghosh et al. 2007,
33=Fabbiano et al. 2003, 34=Pellegrini et al. 2003b.}

\end{table*}

\end{document}